\documentclass[aps,prb,twocolumn,showpacs,citeautoscript,reprint,longbibliography,superscriptaddress]{revtex4-1}
\usepackage{graphicx}
\usepackage{bm,amsmath,amssymb,wasysym,amsfonts,dcolumn}
\usepackage[colorlinks=true, urlcolor=blue, linkcolor=blue, citecolor=blue, pdfborder={0 0 0}]{hyperref}
\usepackage[usenames,dvipsnames]{xcolor}
\usepackage{mathrsfs}

\newcommand{\be}{\begin{equation}}
\newcommand{\ee}{\end{equation}}
\newcommand{\bea}{\begin{eqnarray}}
\newcommand{\eea}{\end{eqnarray}}

\begin{document}

\title{Interface reflectivity of a superdiffusive spin current in ultrafast demagnetization and THz emission}
\author{Wen-Tian Lu}
\author{Yawen Zhao}
\affiliation{Center for Advanced Quantum Studies and Department of Physics, \\ Beijing Normal University, Beijing 100875, China}
\author{Marco Battiato}
\affiliation{School of Physical and Mathematical Sciences, Physics and Applied Physics, Nanyang Technological University, 21 Nanyang Link, Singapore, Singapore}
\author{Yizheng Wu}
\affiliation{Department of Physics, State Key Laboratory of Surface Physics, Fudan University, Shanghai 200433, China}
\affiliation{Collaborative Innovation Center of Advanced Microstructures, Nanjing 210093, China}
\author{Zhe Yuan}
\email[Corresponding author: ]{zyuan@bnu.edu.cn}
\affiliation{Center for Advanced Quantum Studies and Department of Physics, \\ Beijing Normal University, Beijing 100875, China}

\date{\today}

\begin{abstract}
The spin- and energy-dependent interface reflectivity of a ferromagnetic (FM) film in contact with a nonmagnetic (NM) film is calculated using a first-principles transport method and incorporated into the superdiffusive spin transport model to study the femtosecond laser-induced ultrafast demagnetization of Fe$|$NM and Ni$|$NM (NM= Au, Al \& Pt) bilayers. By comparing the calculated demagnetization with transparent and real interfaces, we demonstrate that the spin-dependent reflection of hot electrons has a noticeable influence on the ultrafast demagnetization and the associated terahertz electromagnetic radiation. In particular, a spin filtering effect is found at the Fe$|$NM interface that increases the spin current injected into the NM metal, which enhances both the resulting demagnetization and the resulting THz emission. This suggests that the THz radiation can be optimized by tailoring the interface, indicating a very large tunability.
\end{abstract}

\maketitle

\section{\label{sec:level1}Introduction}\label{chap1}

In spintronics, the transport properties of electrons can be controlled using magnetic configurations, which determine the local potentials of conduction electrons depending on their spins. \cite{Grunberg1988,Fert1988,wolf2001spintronics} The reciprocal process suggests that the local magnetization can be manipulated by a spin-dependent charge current or a pure spin current via the so-called spin-transfer torque.\cite{berger1996emission,slonczewski1996current} The magnetization dynamics induced by a spin-transfer torque or an external magnetic field are usually on the time scale of nanoseconds or in the gigahertz regime, which has been well described by the phenomenological Landau-Lifshitz-Gilbert equation \cite{gilbert1955lagrangian} in the past half century.

The above understanding of magnetization dynamics was challenged by the discovery of the laser-induced magnetization variation in the ferromagnetic metal Ni within hundreds of femtoseconds, which was first observed in a time resolved magneto-optical Kerr effect experiment. \cite{beaurepaire1996ultrafast} The ultrafast demagnetization was later confirmed via other experimental techniques, such as second harmonic generation,\cite{hohlfeld1997nonequilibrium} and two-photon photoemission.\cite{scholl1997ultrafast} One of the key puzzles in this phenomenon is the unknown mechanism that dissipates the spin angular momentum on such a short time scale, \cite{regensburger2000time, koopmans2000ultrafast, oppeneer2004ultrafast} which has stimulated more experimental studies. By means of element-resolved x-ray magnetic circular dichroism, Stamm {\it et al.} unambiguously demonstrated that electron orbitals were not the dissipation channel for spin angular momentum. \cite{stamm2007femtosecond} A spin-polarized charge current was observed during the ultrafast demagnetization, indicating that the spin-dependent transport provided the ultrafast mechanism to transfer spin angular momentum. \cite{malinowski2008control,melnikov2011ultrafast}

\begin{figure}[b]
  \centering
  \includegraphics[width=\columnwidth]{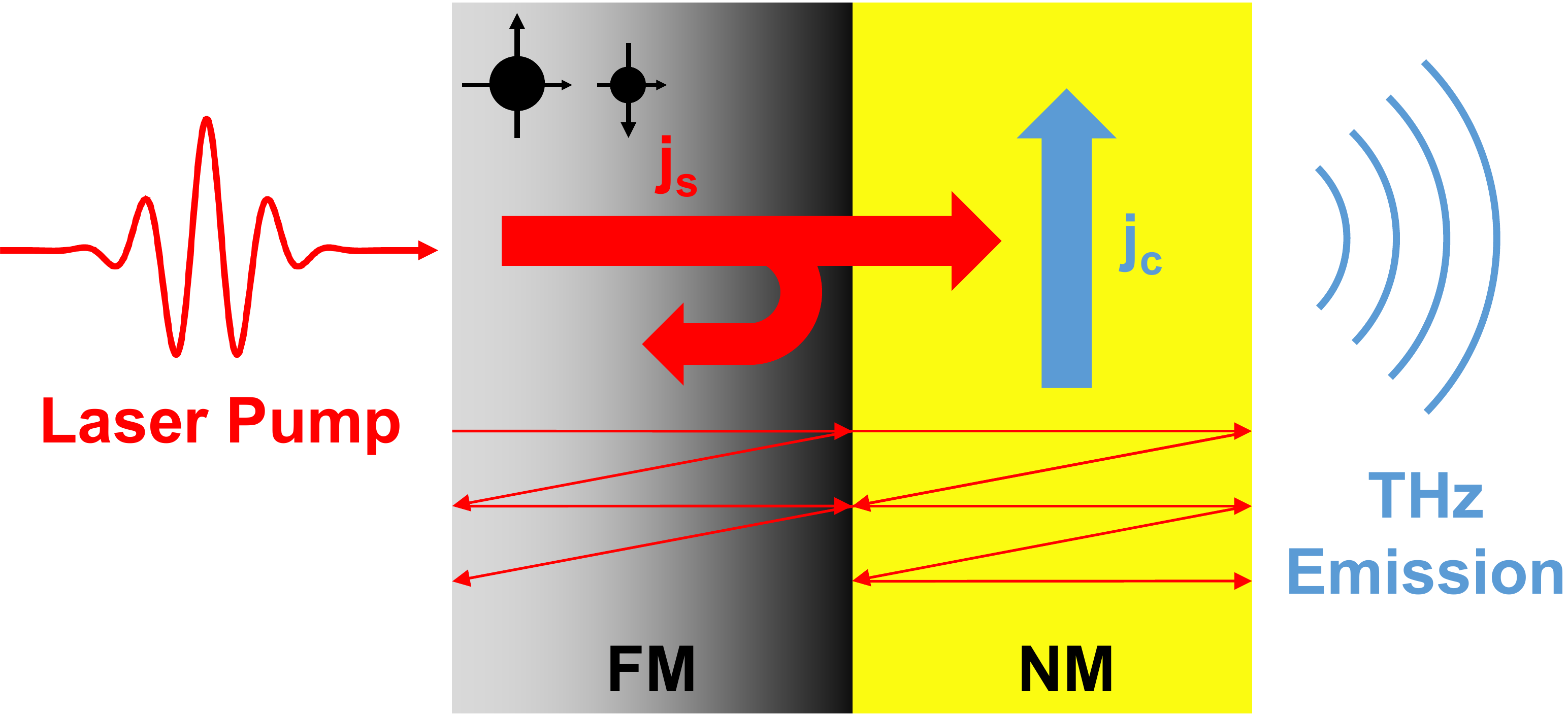}
  \caption{Schematic illustration of ultrafast spin transport and THz emission in an FM$|$NM bilayer. The hot electrons in the FM metal are excited by a laser pulse. At the interface, these hot electrons are partly reflected, and the reflectivity depends on the spin of the electrons. The resulting spin-polarized current entering the NM layer is converted to a transverse charge current via the inverse spin Hall effect. The transverse charge current pulse generates  electromagnetic emission, which is in the THz frequency regime. }
  \label{FMNM}
\end{figure}
In the past two decades, much effort has been made to improve the theoretical understanding and description of the ultrafast demagnetization. Beaurepaire {\it et al.} \cite{beaurepaire1996ultrafast} first applied a phenomenological three-temperature model that takes into account the interactions of the electron, spin, and lattice. Later, many physical mechanisms based on local spin flip were proposed to explain the ultrafast dissipation of spin angular momentum. \cite{koopmans2005unifying, carpene2008dynamics, krauss2009ultrafast, bigot2009coherent, lefkidis2009angular, koopmans2010explaining, zhang2000laser, fahnle2011electron} More recently, spin-dependent hot-electron transport was proposed to be the main mechanism for ultrafast dissipation of spin angular momentum. \cite{battiato2010superdiffusive} These models, such as the superdiffusive spin transport model \cite{battiato2010superdiffusive,battiato2012theory} and the Boltzmann transport theory based on particle-in-cell simulation, \cite{nenno2018particle} capture the main physical process of the laser-induced ultrafast demagnetization.

If the FM metal is attached to an NM metal, then the laser-excited hot electrons can transport across the interface and enter the NM metal. The resulting femtosecond pulse of a spin-polarized charge current in the NM metal can be converted to a transverse charge current via the inverse spin Hall effect, \cite{valenzuela2006direct, saitoh2006conversion} which in turn generates terahertz electromagnetic emission;\cite{kampfrath2013terahertz} see Fig.~\ref{FMNM}. The magnetic multilayers become an effective THz source with the advantages of structural compactness, low cost, flexibility and broadband. \cite{seifert2016efficient,feng2018highly,herapath2019impact,nenno2019modification}

In the current theory of superdiffusive spin transport, which is applied in the description of both ultrafast demagnetization and THz emission, the FM$|$NM interface is usually assumed to be practically transparent, indicating that all electrons can freely pass through the interface from either side. In reality, the laser-excited hot electrons experience different lattice potentials on the two sides of the interface and therefore are partly reflected, resulting in the so-called interface resistance \cite{schep1997interface}. Unlike the conventional interface resistance, which is a measure of the reflectivity of conduction electrons at the Fermi level, one needs to consider the reflectivity of the hot electrons at energies higher than the Fermi energy. The interface reflection of an ultrafast spin current was proposed and formulated in detail, \cite{battiato2014treating} and model transmission of electrons across an interface was studied. \cite{rudolf2012ultrafast,kampfrath2013terahertz} Without a quantitative estimation of the energy- and spin-dependent reflectivity of real interfaces, it is difficult to examine the influence of interface reflection on ultrafast demagnetization and terahertz emission based on the theoretical superdiffusive spin transport model. \cite{rudolf2012ultrafast, eschenlohr2013ultrafast, kampfrath2013terahertz}

In this article, we extend superdiffusive spin transport theory to take into account the real reflectivity at the FM$|$NM interface, which is calculated using a first-principles transport method based on the local spin density approximation of density-functional theory. The calculated spin- and energy-dependent reflectivities are then incorporated into the calculation of the superdiffusive spin current. \cite{battiato2014treating} Using Au, Al, and Pt as typical examples of an NM metal, we explicitly calculate the laser-induced ultrafast demagnetization and THz emission for Fe$|$NM and Ni$|$NM bilayers. The difference in the calculated results between the transparent and real interface is systematically illustrated. In particular, the Fe$|$NM interface exhibits a lower reflectivity for the spin-up electrons than for the spin-down electrons. Such spin filtering effects at the interface effectively increase the spin current injected into the NM metal and therefore enhance the demagnetization and THz emission. The methods presented in this paper can be straightforwardly applied in studying multilayer structures with multiple interfaces, such as spin valves. \cite{balavz2018transport} The rest of this paper is organized as follows. The first-principles calculation of the interfacial reflectivity is presented in Sec.~\ref{chap2}. The generalized superdiffusive spin transport theory is given in Sec.~\ref{chap3}, followed by the calculated ultrafast demagnetization of the Fe$|$NM and Ni$|$NM bilayers. Section~\ref{chap4} shows the THz emission calculated using the real reflectivity of the FM$|$NM interface. The conclusions are drawn in Sec.~\ref{chap5}.

\section{\label{sec:level2}Spin- and Energy-Dependent Interface Reflectivity}\label{chap2}

\subsection{Theoretical Methods and Computational Details}

With the aim of obtaining the spin- and energy-dependent reflectivity due to interface scattering, we employ a well-developed transport theory based on the Landauer-B{\"u}ttiker formalism combined with the detailed electronic structure of the real materials obtained from a first-principles calculation \cite{xia2001interface}. Without introducing any free parameters, this method can be used to calculate the finite-temperature resistivity of transition metals \cite{Liu:prb15} and the interface resistance between two metals \cite{Xia:prb06}, in good agreement with experiments. The transport calculation is generalized from the Fermi level for quasi-equilibrium states to higher energies to describe the electron and/or spin transport due to excitation by a laser pulse. We present the theoretical and numerical details of our calculation in this section.

First, we define the terminology used in this paper. A ``transparent'' interface refers to the interface with zero reflectivity. For the real interfaces, whose reflectivity is calculated using first-principles method, we consider clean and disordered interfaces, respectively. The ``clean'' interface is modeled with a sharp planar boundary between two materials while the ``disordered'' one has atomic interdiffusion resulting in thin alloying layers at the interface.

\begin{figure}[b]
  \centering
  \includegraphics[width=\columnwidth]{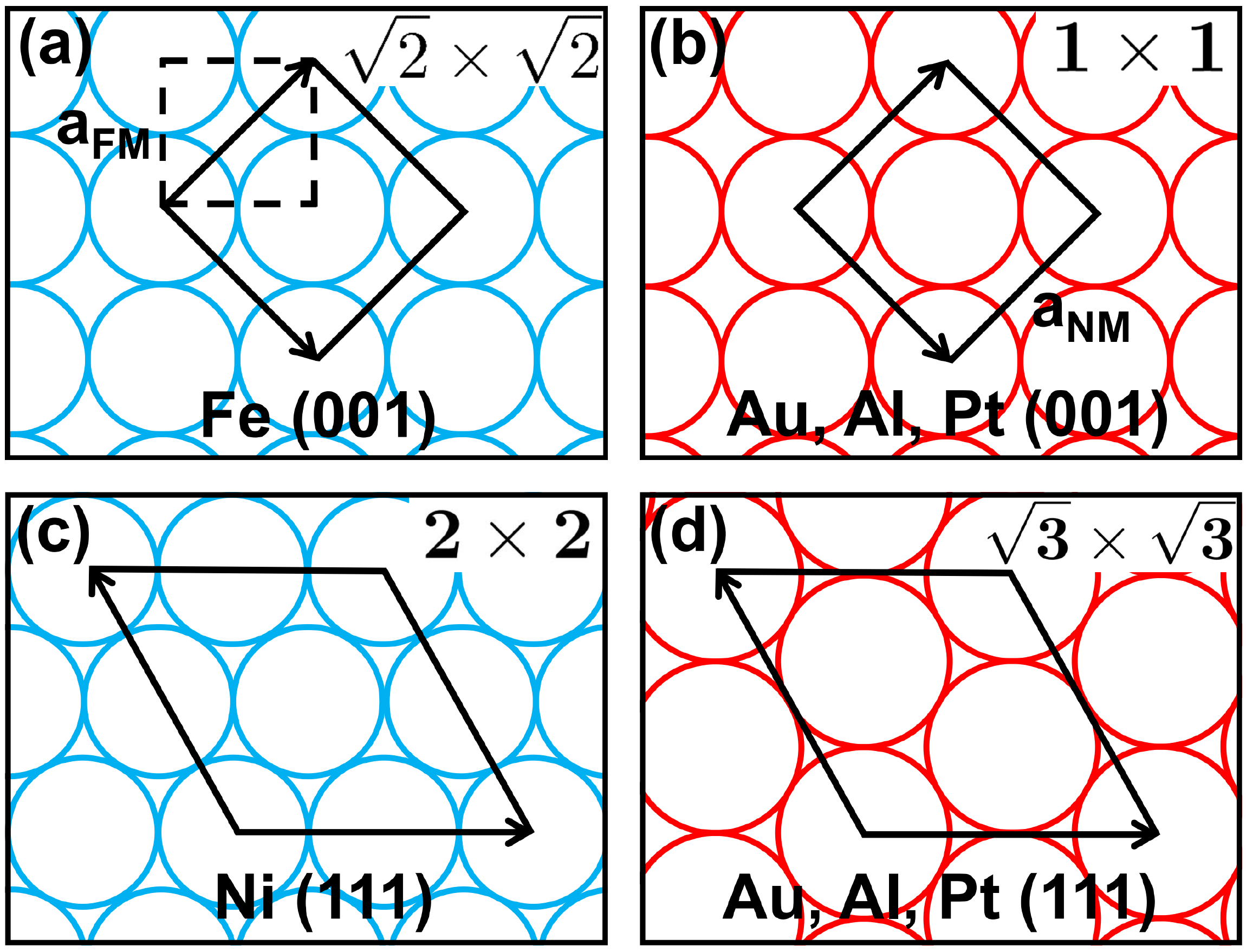}
  \caption{Two-dimensional (2D) unit cell of the Fe$|$NM(001) and Ni$|$NM(111) interface in the atomic plane perpendicular to the layer stacking direction.}
  \label{fig:match}
\end{figure}

To model the interface between FM (Fe and Ni) and NM metals (Au, Al, and Pt) whose lattices are not matched, we construct periodic supercells in the directions transverse to the transport direction, as schematically illustrated in Fig.~\ref{fig:match}. The lattice constants of face-centered cubic (FCC) Au ($a_{\rm Au}=4.050$~{\AA}) and Al ($a_{\rm Al}=4.090$~{\AA}) are approximately equal to $\sqrt{2}a_{\rm Fe}$, where $a_{\rm Fe}=2.866$~{\AA} is the lattice constant of body-centered cubic (BCC) Fe. Thus, the two-dimensional (2D) $1\times 1$ Au(001) and Al(001) unit cells are perfectly matched with the $\sqrt{2}\times\sqrt{2}$ supercell of Fe(001), as plotted in Fig.~\ref{fig:match}(a) and (b). The lattice constant of FCC Pt is slightly smaller than $\sqrt{2}a_{\rm Fe}$, and thus, we either stretch Pt or compress Fe to make a lattice-matched interface. In practice, the calculated interface reflectivity varies by less than 10\% between the two cases. Analogously, we use a $2\times2$ Ni(111) supercell to match the $\sqrt{3}\times\sqrt{3}$ NM(111) cell and construct the Ni$|$NM(111) interfaces, as shown in Fig.~\ref{fig:match}(c) and (d). Unless stated otherwise, all the results reported in this paper are carried out with the lattice constants of $a_{\rm Fe}=2.866$~{\AA} and $a_{\rm Ni}=3.524$~{\AA} unchanged.

The constructed interfaces are connected to the semi-infinite FM and NM metals on both sides. Within the framework of density functional theory, the electronic structure of the interface is self-consistently determined in the atomic-sphere approximation using the surface Green's function method \cite{turek2013electronic} implemented with a tight-binding linear muffin-tin orbital (TB-LMTO) basis \cite{Andersen1985}. A minimal basis consisting of $s$, $p$ and $d$ orbitals is used, and the 2D Brillouin zone is sampled by a $\mathbf k_\|$ mesh with a constant density corresponding to $120\times120$ for a $1\times1$ FM unit cell.

Having the self-consistent atomic potentials of such interfaces, we solve the quantum scattering problem using a wave-function matching method, which is also implemented using the TB-LMTO basis. Note that the interface resistance arises mainly from the different potentials experienced by conduction electrons across the interface, \cite{kritithesis} which is of the order of the exchange splitting energy and/or the difference in the work functions of the two metals. Such an energy scale is much larger than that of spin-orbit coupling. Therefore, spin-orbit coupling can be neglected and for each spin channel $\sigma$, all of the propagating Bloch states at every $\mathbf k_\|$ and energy $E$ in the FM or NM lead are explicitly calculated. The Sharvin conductance of the material is then determined by the total number of propagating states $N^\sigma(\mathbf k_\|, E)$,
\begin{equation}
G_{\rm Sh}^{\sigma}(E)=\frac{e^2}{h A N_{k_\|}}\sum_{\mathbf k_\|}N^{\sigma}(\mathbf k_\|,E).
\end{equation}
Here, $e^2/h$ is the quantized electronic conductance for a single spin channel, $N_{k_\|}$ is the total number of $\mathbf k_\|$ in the 2D Brillouin zone and $A$ is the cross-sectional area of the lateral supercell in the calculation. Note that the NM metal has equal Sharvin conductances for both spin channels, i.e., $G_{\rm Sh, NM}^{\uparrow}(E)=G_{\rm Sh, NM}^{\downarrow}(E)$.

Then, the spin-dependent transmission probability amplitude $t^{\sigma}_{\mu\nu}(\mathbf k_\|,E)$ from the $\nu$-th propagating state incoming from the left lead to the $\mu$-th outgoing state in the right lead is evaluated at energy $E$ such that the interface conductance is \cite{datta1997electronic}
\be
G^{\sigma}(E)=\frac{e^2}{h A N_{k_\|}} \sum_{\mathbf k_\|} \sum_{\mu,\nu} \left\vert t^{\sigma}_{\mu\nu}(\mathbf k_\|,E)\right\vert^2.
\ee
Then, we define the spin-dependent interface reflectivity from the FM and NM sides, respectively, as \cite{Gijs1997perpendicular}
\be
R^{\sigma}_{\mathrm{FM} \rightarrow \mathrm{NM}}(E) =1 -\frac{G^{\sigma}(E)}{G_{\rm{Sh, FM}}^{\sigma}(E)}
\label{eqnreflect}
\ee
and
\be
R^{\sigma}_{\mathrm{FM} \leftarrow \mathrm{NM}}(E) =1 -\frac{G^{\sigma}(E)}{G_{\rm{Sh, NM}}^{\sigma}(E)}.
\label{eqnreflectNM}
\ee
Note that although the conductance is invariant for electrons incoming from either side of the interface, the reflectivity can be different owing to the different Sharvin conductances of the FM and NM metals. In addition, for the study of the superdiffusive electron and/or spin transport excited by a laser pulse, we consider the energy range from the Fermi level to 1.5~eV above the Fermi level.

At the metallic interface, the interdiffusion of atoms may result in an interfacial alloy. To model such disordered interfaces, we build a large lateral supercell and randomly distribute two types of atoms in certain interfacial atomic layers consisting of 50\%-50\% alloy. To ensure accuracy, we take 10 random configurations of the disordered interface for each energy and then average the results. Generally, we apply two atomic layers of the alloy at the disordered interface, and the influence of the thickness of the disordered interfacial layers on the interfacial reflectivity is explicitly examined for the Fe$|$Au bilayer.

\subsection{Fe$|$NM Interface}

\begin{figure}[t]
  \centering
   \includegraphics[width=0.8\columnwidth]{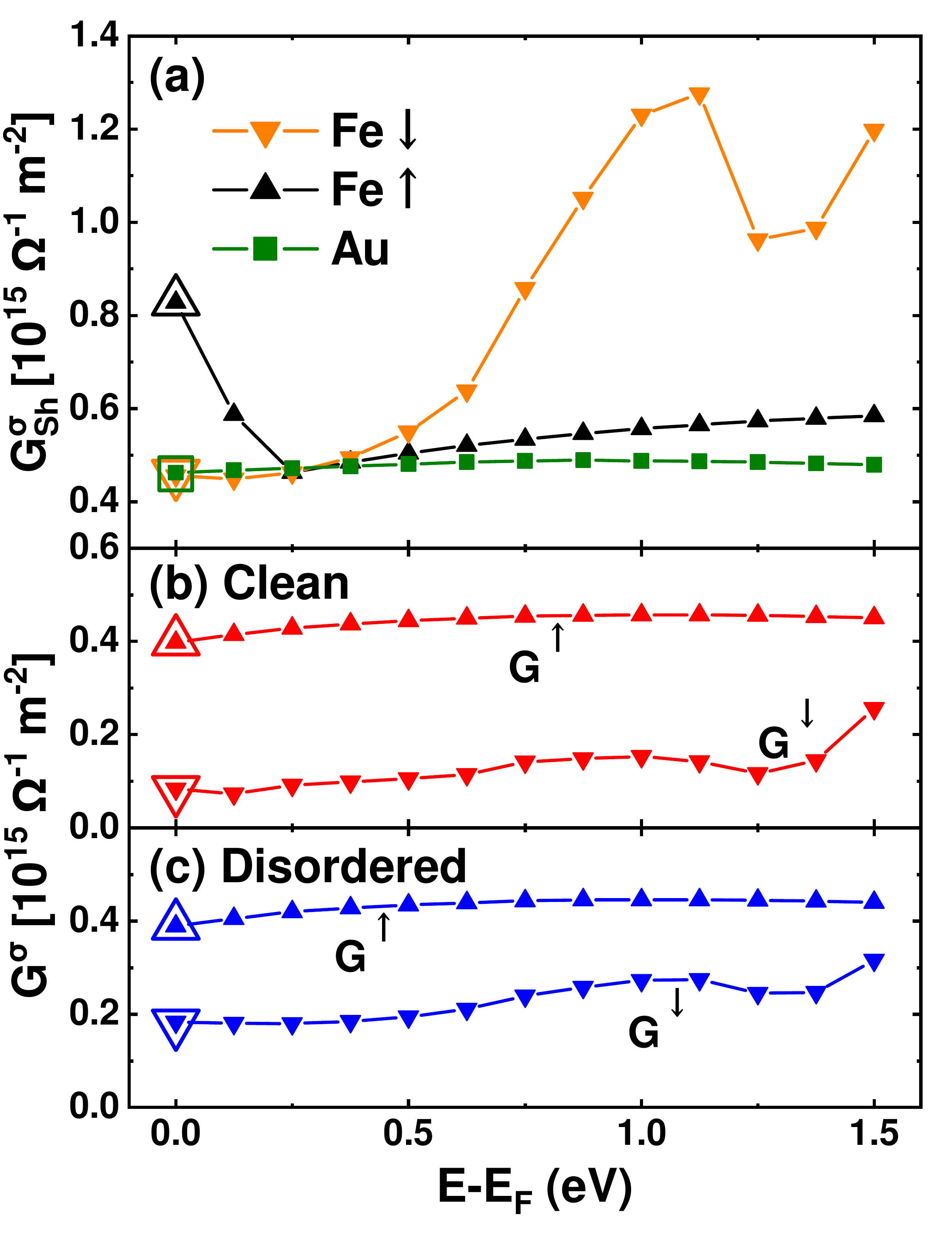}
  \caption{(a) Calculated Sharvin conductances of Fe and Au as a function of energy. (b) Calculated spin-dependent conductance of a clean Fe$|$Au interface. (c) Calculated spin-dependent conductance of a disordered Fe$|$Au interface, where Fe and Au atoms are mixed with equal concentration in the two atomic layers at the interface. The large empty symbols represent the calculated values at the Fermi level in the literature. \cite{zwierzycki2005first}
  }
  \label{fig:FeAu-conductance}
\end{figure}
Figure~\ref{fig:FeAu-conductance}(a) shows the calculated Sharvin conductances of Fe and Au as functions of energy. In particular, the Sharvin conductances at the Fermi level are in perfect agreement with those in the literature \cite{zwierzycki2005first}, which are plotted as large empty symbols for comparison. For Au, the Sharvin conductance exhibits very little energy dependence and is nearly constant ($0.5\times10^{15}~\Omega^{-1}\,\text{m}^{-2}$). In contrast, a significant energy dependence is found for the Sharvin conductance of Fe. As the energy increases, $G^{\uparrow}_{\rm Sh}$ first decreases from $0.83\times10^{15}~\Omega^{-1}\,\text{m}^{-2}$ at the Fermi level to $0.46\times10^{15}~\Omega^{-1}\,\text{m}^{-2}$ at an energy of 0.25~eV above $E_F$. This occurs because the number of propagating states becomes smaller at energies above the top of the 3$d$ spin-up bands. This decrease is then followed by a slight increase, where $G^{\uparrow}_{\rm Sh}$ is mostly contributed to by the $s$ electrons. For the spin-down electrons, $G^{\downarrow}_{\rm Sh}$ shows a remarkable increase at 0.5~eV above $E_F$ due to the unoccupied 3$d$ bands.

The calculated conductances for clean and disordered Fe$|$Au interfaces are plotted in Fig.~\ref{fig:FeAu-conductance}(b) and (c). The spin-up conductances of the clean and disordered interfaces are both approximately $0.4\times10^{15}~\Omega^{-1}$\,m$^{-2}$ and are independent of the energy, indicating the small effect of the interface disorder on the spin-up conductance. The calculated conductance of the spin-down electrons, which depends slightly on the energy, is increased by the interface disorder. This occurs because the spin-down conducting channels in Fe do not match those of Au in the 2D reciprocal space. Nevertheless, the interface disorder breaks the conservation of momentum and hence increases electron transmission through the interface. \cite{xia2001interface}

\begin{figure}[t]
  \centering
  \includegraphics[width=\columnwidth]{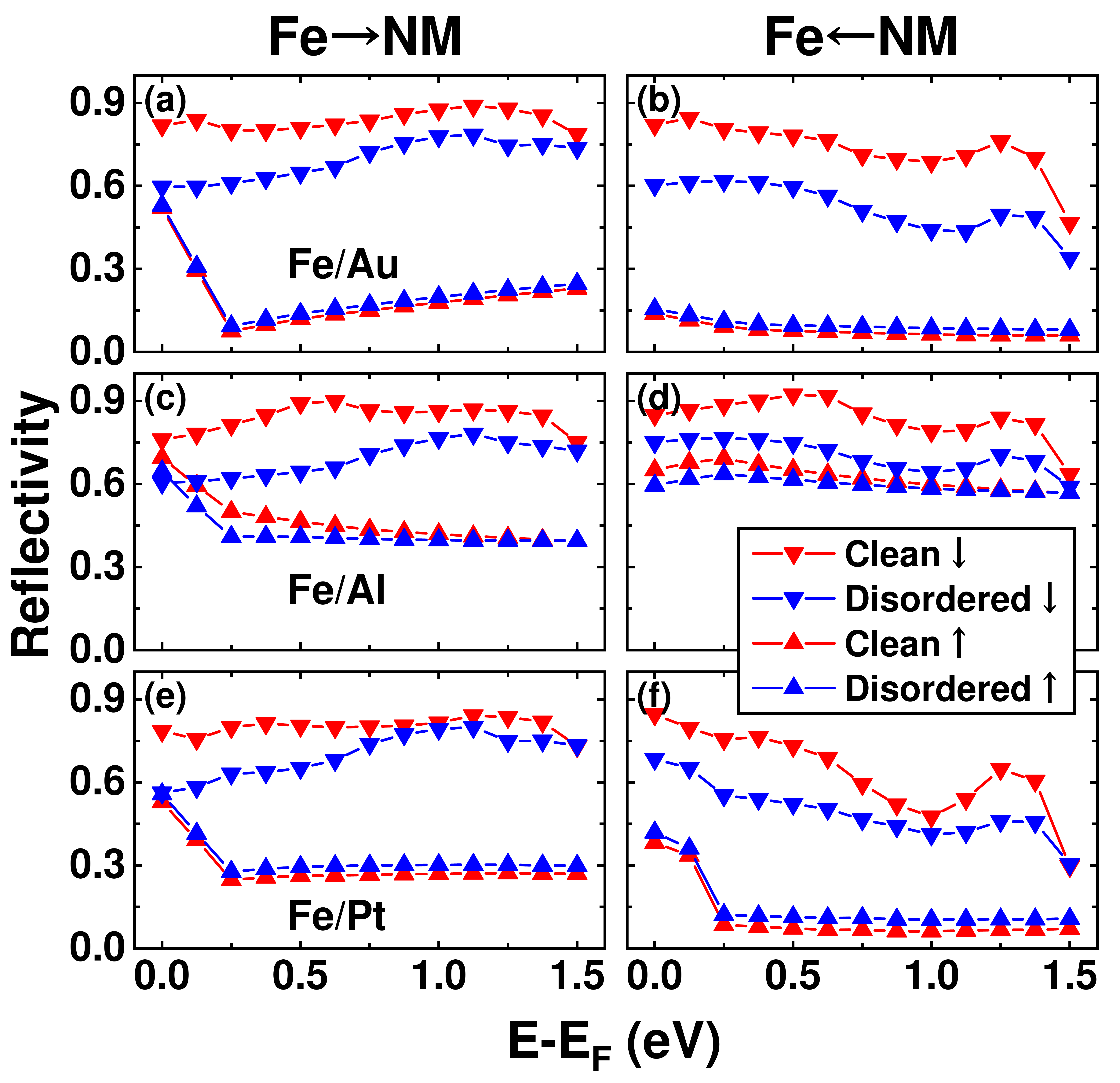}
  \caption{Calculated reflectivity of Fe$|$NM(001) interfaces. Panels on the left-hand (right-hand) side represent the Fe$\rightarrow$NM (Fe$\leftarrow$NM) reflectivities for NM=Au [(a) and (b)], Al [(c) and (d)] and Pt [(e) and (f)]. The up and down triangles denote the spin-up and spin-down channels, respectively. The reflectivities for clean interfaces are plotted as red symbols, while those for disordered interfaces are shown as blue symbols.}
  \label{fig:R-FeNM}
\end{figure}
The reflectivity for the Fe$|$Au(001) interface can be calculated using Eq.~\eqref{eqnreflect} and Eq.~\eqref{eqnreflectNM}, as shown in Fig.~\ref{fig:R-FeNM}(a) and (b), respectively. For the direction of Fe$\rightarrow$Au, the reflectivities of the spin-up electrons are nearly identical for clean and disordered interfaces. $R^\uparrow_{\mathrm{Fe}\rightarrow\mathrm{Au}}$ decreases when the energy goes from $E_F$ to $E_F+0.25$~eV and slightly increases at higher energy. The energy dependence mainly arises from the energy-dependent Sharvin conductance of Fe $G^{\uparrow}_{\rm Sh, Fe}$. Compared with the spin-up reflectivity, $R^\downarrow_{\mathrm{Fe}\rightarrow\mathrm{Au}}$ is larger due to the small interface conductance and exhibits a weaker variation as a function of energy. Therefore, the spin-up electrons will pass more easily through the interface from Fe to Au than the spin-down electrons, which is qualitatively consistent with recent calculations. \cite{alekhin2017femtosecond} The difference in the reflectivity between the two spin channels results in spin filtering for the electrons passing through the Fe$|$Au(001) interface.

The calculated reflectivity from Au to Fe is plotted in Fig.~\ref{fig:R-FeNM}(b). For spin-up electrons, the reflectivities for clean and disordered interfaces are approximately as small as  0.1. They have minimal energy dependence because both the Sharvin conductance of Au and the spin-up conductance of the Fe$|$Au(001) interface do not depend on the energy. For the spin-down electrons, a slight increase in $G^{\downarrow}$ leads to a decrease in the reflectivity, which is particularly prominent near 1.5~eV. The larger $G^{\downarrow}$ for the disordered interface results in a lower reflectivity than in the clean case.

\begin{figure}[t]
  \centering
  \includegraphics[width=\columnwidth]{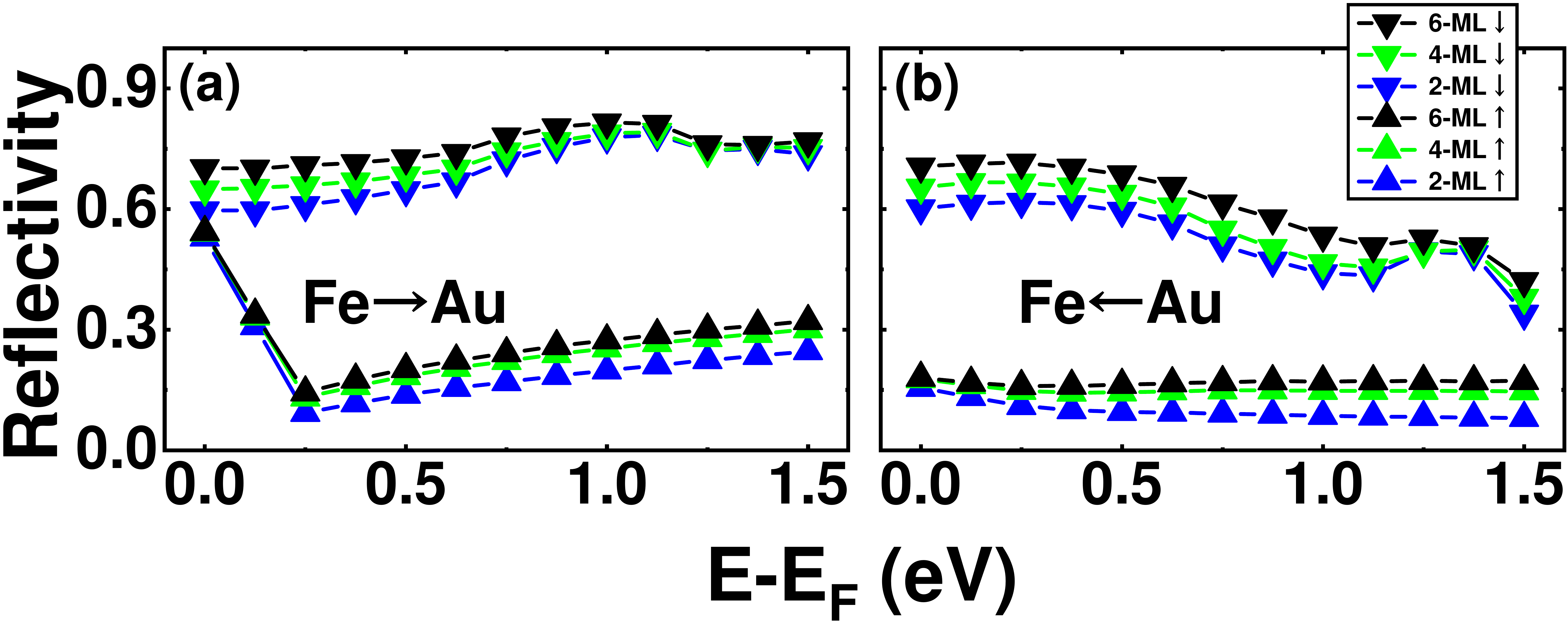}
  \caption{Calculated reflectivity of the Fe$|$Au(001) interface as a function of energy for different thicknesses of the interfacial alloy.}
  \label{fig:R-FeAu}
\end{figure}
At the real interface, little is known about how deep Fe atoms may diffuse into the Au layer or how thick the interfacial alloy can become. Thus, we model the interfacial alloy at the Fe$|$Au(001) interface with two, four and six atomic layers, where the Fe and Au atoms are randomly distributed with equal probability. Here, we consider the same scheme to deal with the different atomic sizes of Fe and Au as in the previous calculation, \cite{zwierzycki2005first} where the diffused atoms are supposed to occupy atomic spheres of the same size as that of the host element. Then, we repeat the calculation of the interfacial reflectivity, which is plotted in Fig.~\ref{fig:R-FeAu}. When the alloy thickness increases from two to six atomic layers, the reflectivity only slightly increases for all the cases, and the reflectivity does not show much dependence on the thickness of the interfacial alloy. Therefore, we use two atomic layers of the alloy to simulate the interface disorder in the following sections of this paper.

The calculated reflectivities of Fe$|$Al(001) are displayed in Fig.~\ref{fig:R-FeNM}(c) and (d). Similar to the case of Fe$\rightarrow$Au, the spin-down reflectivity from Fe to Al through a clean interface is approximately 0.8 and is slightly lowered by interface disorder.  The spin-up reflectivity for Fe$\rightarrow$Al is nearly constant (0.4) from 0.3 eV to 1.5 eV above the Fermi level. Near $E_F$, the spin-up reflectivity decreases with increasing energy owing to the energy-dependent $G^{\uparrow}_{\rm Sh,Fe}$. In the Al$\leftarrow$Fe direction shown in Fig.~\ref{fig:R-FeNM}(d), the spin-up reflectivity is approximately 0.6 for the whole energy range, which is not influenced by interface disorder. The large reflectivity arises from the large Sharvin conductance of Al. The spin-down reflectivities from both sides of Fe$|$Al are comparable and are slightly reduced by interface disorder.

For the Fe$|$Pt interface, the calculated reflectivity shown in Fig.~\ref{fig:R-FeNM}(e) and (f) shares the main features of Fe$|$Au(001). The spin-up reflectivities from both sides are significantly lower than the spin-down reflectivities. In addition, interface disorder slightly lowers the spin-down reflectivity but has little effect on the spin-up reflectivity. The spin-up reflectivities from both sides increase with decreasing energy near the Fermi level, while the spin-down reflectivity from Pt to Fe exhibits a drop at $E_F+1.5$~eV.

The calculated spin-down reflectivities for all three cases of Fe$|$NM interfaces shown in Fig.~\ref{fig:R-FeNM} are always larger than the reflectivity of the spin-up electrons, independent of the transport direction. This common feature indicates that the Fe$|$NM interface plays the role of a spin filter, which improves the spin angular momentum transfer from the Fe layer to the NM layer. We can understand this effect from the electronic structure of the ferromagnetic Fe, whose 3$d$ bands of majority spin are mostly filled and whose 4$s$ electrons dominate the calculated energy range. This can also be seen from the energy independence of the calculated spin-up reflectivity above $E_F+0.2$~eV. The electronic structure of all three NM metals, Au, Al, and Pt, are dominated by the $s$ electrons (except for Pt) near $E_F$. The delocalization of the $s$ electrons improves the spin-up conductance and hence lowers the spin-up reflectivity. For the minority-spin channel, on the other hand, the presence of the unoccupied $d$ electrons of Fe, which are mismatched with the $s$ electrons of the NM metal, markedly increases the electron scattering at the interface, resulting in a large spin-down reflectivity. By allowing the transmission from $\mathbf k_\|$ to $\mathbf k'_\|$, interface disorder increases the spin-down conductance. Therefore, the spin-down reflectivity at the disordered interface is slightly lower than that at the clean interface.

\subsection{Ni$|$NM Interface}

\begin{figure}[b]
  \centering
  \includegraphics[width=\columnwidth]{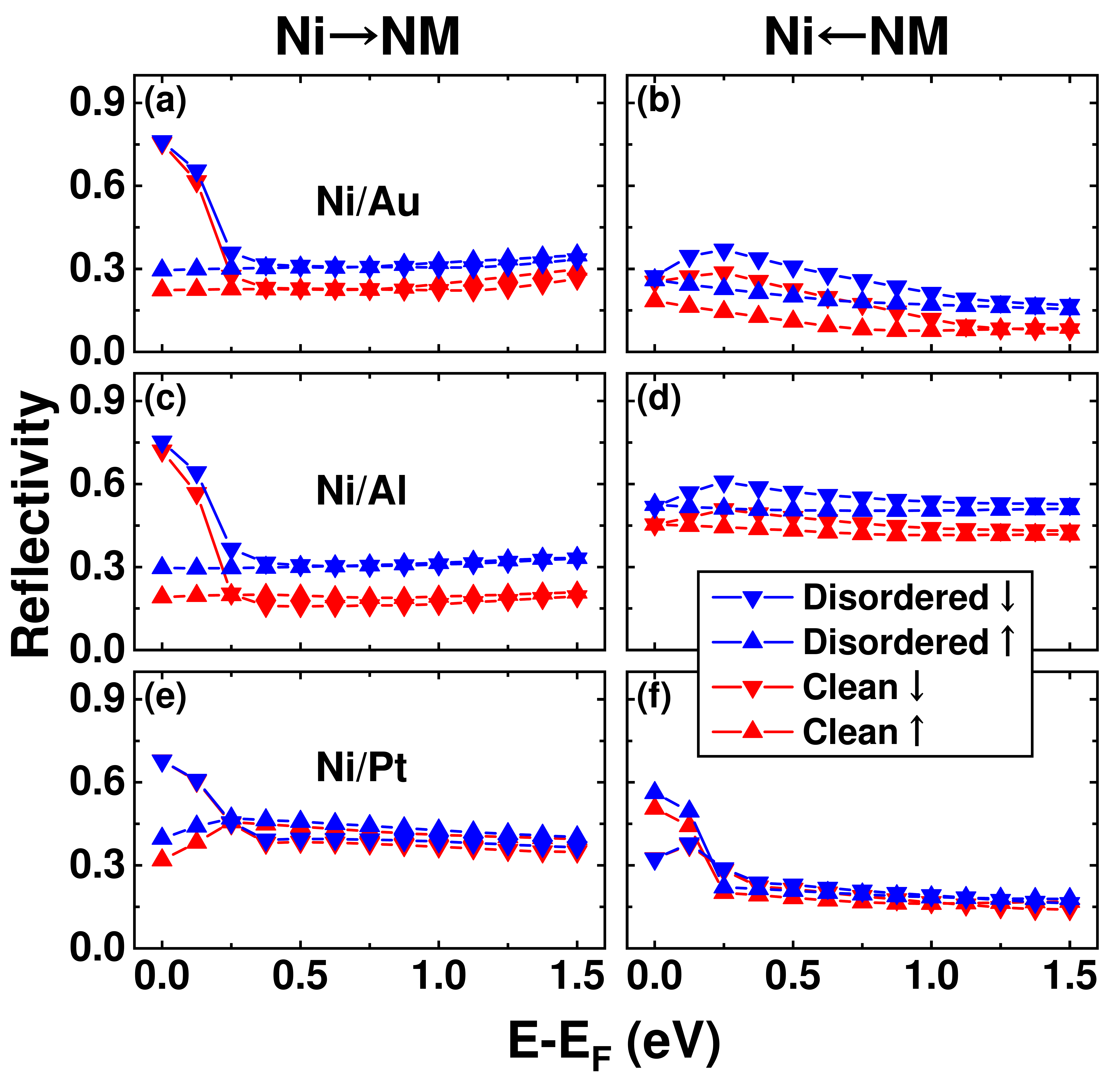}
  \caption{Calculated reflectivity of the Ni$|$NM(111) interface. Panels on the left-hand (right-hand) side represent the Ni$\rightarrow$NM (Ni$\leftarrow$NM) reflectivities for NM=Au [(a) and (b)], Al [(c) and (d)] and Pt [(e) and (f)]. The up and down triangles denote the spin-up and spin-down channels, respectively. The reflectivities for clean interfaces are plotted as red symbols, while those for disordered interfaces are shown as blue symbols.}
  \label{fig:R-NiNM}
\end{figure}
We then calculate the interface reflectivities for Ni$|$NM(111) bilayers with NM=Au, Al, and Pt, as shown in Fig.~\ref{fig:R-NiNM}, where we find the following two distinct features compared to the reflectivities of Fe$|$NM interfaces in Fig.~\ref{fig:R-FeNM}: (1) The spin-up and spin-down reflectivities are nearly degenerate in most of the calculated energy range. There is a sharp rise in the spin-down reflectivity from Ni to the NM metal near the Fermi energy. (2) Interface disorder increases the calculated reflectivity for both spin-up and spin-down electrons. These two properties can be understood in terms of the electronic structure of Ni: unlike Fe, the spin-up 3$d$ bands of Ni are fully occupied, and the spin-down 3$d$ bands are mostly occupied. Thus, the electronic structure is dominated by 4$s$ electrons in the energy range we study, except for 3$d$ spin-down electrons near the Fermi level. The latter are responsible for the high spin-down reflectivity of Ni$\rightarrow$NM near the Fermi level, as seen in Fig.~\ref{fig:R-NiNM}(a), (c) and (e), indicating that a large fraction of 3$d$ spin-down electrons from Ni do not go through the interface. The calculated reflectivity from Pt to Ni, which is shown in Fig.~\ref{fig:R-NiNM}(f), is large near $E_F$ because the 5$d$ electrons are highly reflected at the interface. Because delocalized $s$ electrons dominate the transport on both the Ni and NM sides, mismatch in conduction channels in the 2D Brillouin zone, as at the Fe$|$Au interface, does not exist. Therefore, interface disorder does not improve the electron transmission through the interface. Instead, the disorder scattering due to the interfacial alloy increases the reflectivity such that the blue symbols in Fig.~\ref{fig:R-NiNM} are always higher than the corresponding red ones. Owing to the comparable reflectivities for spin-up and spin-down electrons, the Ni$|$NM interfaces do not act as spin filters, which is significantly different from the Fe$|$NM interfaces.

\section{\label{sec:level3}Ultrafast demagnetization}\label{chap3}

\subsection{The Superdiffusive Spin Transport Model}

Superdiffusive spin transport theory has been successfully applied in describing the ultrafast demagnetization resulting from excitation by a femtosecond laser pulse. In this theory, a laser pulse creates many nonequilibrium electrons, which are excited from the occupied states below the Fermi level to the energy bands above $E_F$. The characteristics of these nonequilibrium electrons, e.g., their group velocities and lifetimes, depend on the magnetic material and the excitation energy. For instance, the majority-spin 3$d$ electrons in Fe are excited to the $s$ band by a laser with energy 1.5~eV, while the minority-spin 3$d$ electrons are excited to other unoccupied 3$d$ bands. Therefore, the excited nonequilibrium electrons in Fe have different group velocities depending on their spins. Meanwhile, holes are left in the 3$d$ bands below $E_F$ and are neglected in the superdiffusive spin transport model because of their relatively low mobility. The nonequilibrium electrons above $E_F$ move because of their finite group velocities until they are scattered by phonons, impurities, or other electrons. If the scattering is elastic, then the electrons do not lose energy; however, their momenta are changed by the scattering. Inelastic scattering allows recombination of nonequilibrium electrons with the holes below the Fermi level, where the released energy may excite other electrons to higher energies, giving rise to a cascade of electrons contributing to the transport. Therefore, the electron transport exhibits a ballistic characteristic right after excitation by a laser pulse and gradually approaches the diffusive regime.

The laser-excited superdiffusive spin current can transport across the FM$|$NM interface, which transports angular momentum out of the magnetic material. This outgoing spin current accelerates the ultrafast demagnetization induced by a laser pulse. The spin current injected into the NM heavy metal generates a transverse charge current via the so-called inverse spin-Hall effect (ISHE)~\cite{Hoffmann:2013,Niimi:2015,Sinova:2015} and then induces THz emission owing to the sub-picosecond time scale of the charge current. Therefore, the FM$|$NM interface plays an important role in the ultrafast demagnetization and THz emission. In the previous superdiffusive transport model, this interface effect was not systematically examined owing to the lack of reliable electron transmission or reflection data. Here, we include our calculated energy- and spin-dependent electron reflectivity in the superdiffusive transport theory following the treatment in Refs.~\onlinecite{battiato2010superdiffusive,battiato2012theory,battiato2014treating}. The main theory is briefly outlined as follows.

Since the laser spot is much larger than the mean free path of the excited hot electrons, this model can be reduced to one dimension, where only the spatial dependence along the interface normal of the metallic multilayers ($z$-axis) is taken into account. The key equation of the superdiffusive spin transport model reads\cite{battiato2010superdiffusive}
\be
\frac{\partial n_{\sigma}(E,z,t)}{\partial t} + \frac{n_{\sigma}(E,z,t)}{\tau_{\sigma}(E,z)} = \left( -\frac{\partial}{\partial z} \hat{\phi}+\hat{I} \right) \times S^{\rm eff}_{\sigma}(E,z,t),\label{eq:master}
\ee
where $n_{\sigma}(E,z,t)$ is the laser-excited hot electron density with spin $\sigma$ at energy $E$, position $z$ and time $t$. $\tau_{\sigma}(E,z)$ is the lifetime of hot electrons with spin $\sigma$ at energy $E$, and it depends on the material at position $z$. $\hat{\phi}$ and $\hat{I}$ in Eq.~\eqref{eq:master} are the electron flux and identity operators, respectively. $S_{\sigma}^{\text{eff}}(E,z,t)$ is the effective source term for the hot electrons including scattered and newly excited electrons. The superdiffusive spin transport equation~\eqref{eq:master} is non-local in space and time. To solve this integro-differential equation, we adopt a recently developed numerical scheme~\cite{battiato2014treating} and discretize space and time into uniform grids. Explicitly, with the discretized time step $\delta t$, the effective source is written as~\cite{battiato2012theory}
\begin{widetext}
\begin{equation}
S^{\rm eff}_{\sigma} \left(E,z,t+\delta t\right) =\sum_{\sigma'} \int dE'\,p_{\sigma,\sigma'}\left(E,E',z\right) \left[1 - e^{-\frac{\delta t}{\tau_{\sigma'}(E',z)}}\right] n_{\sigma'}\left(E',z,t\right) + S^{\rm ext}_{\sigma}\left(E,z,t+\delta t\right).\label{eq:seff}
\end{equation}
Here, $p_{\sigma,\sigma'}\left(E,E',z\right)$ is the transition probability for a hot electron from the state $\{\sigma', E'\}$ to the state $\{\sigma, E\}$ due to scattering, and $S^{\text{ext}}_{\sigma}(E,z,t)$ is the external source of hot electrons that are directly excited by the laser pulse.

The electron flux operator $\hat{\phi}$ acting on a source term can be explicitly written as
\be
\hat{\phi} S(z,t) = \int^{+\infty}_{-\infty} dz_{0} \int^{t}_{-\infty} dt_{0}\ S(z_0,t_0)\phi(z,t|z_0,t_0),
\ee
where the spin and energy indices are omitted for simplicity. The flux kernel $\phi(z,t|z_0,t_0)$ represents the particle density flux at a given point in space and time resulting from an electron that is excited at $z_0$ and $t_0$. Specifically, one has
\be
\phi(z,t|z_0,t_0) = \frac{\widetilde{[\Delta t]}\ e^{-(t-t_0)-\frac{\widetilde{[\frac{\Delta t}{\tau}}]}{\widetilde{[\Delta t]}}}}{2(t-t_0)^2} \Theta\left[t-t_0-\left\vert\widetilde{[\Delta t]}\right\vert\right],\label{eq:kernel}
\ee
where $\Theta$ is the Heaviside step function and the $\Delta$ functions are defined as~\cite{battiato2012theory}
\be
\widetilde{[\Delta t]} (z|z_0) = \int^{z}_{z_0} \frac{dz'}{v(z')} \label{d1},
\ee
and
\be
\widetilde{[\frac{\Delta t}{\tau}]}(z|z_0) = \int^{z}_{z_0} \frac{dz'}{\tau(z')v(z')}.
\label{d2}
\ee
In the above equations, $v(z)$ is the group velocity of hot electrons, which in general depends on the spin $\sigma$, the energy $E$ and the material via the position $z$.

The evolution of the spin density in equation~\eqref{eq:master} is then computed on a discrete spatial grid. The position-dependent reflectivity $R_{\rightleftarrows}(z_i\pm\delta z/2)$ can be explicitly included in the evolution as
\begin{eqnarray}
n(z_i,t+\delta t) &=& n(z_i,t)e^{-\frac{\delta t}{\tau (z_i)}} +S^{\text{eff}}(z_i,t+\delta t)\nonumber\\
&&+\left[1-R_{\rightarrow}\left(z_i-\frac{\delta z}{2}\right)\right]\Phi_{\rightarrow}\left(z_i-\frac{\delta z}{2},t\right)-\left[1-R_{\leftarrow}\left(z_i-\frac{\delta z}{2}\right)\right]\Phi_{\leftarrow}\left(z_i-\frac{\delta z}{2},t\right)\nonumber\\
&&-\left[1-R_{\rightarrow}\left(z_i+\frac{\delta z}{2}\right)\right]\Phi_{\rightarrow}\left(z_i+\frac{\delta z}{2},t\right)+\left[1-R_{\leftarrow}\left(z_i+\frac{\delta z}{2}\right)\right]\Phi_{\leftarrow}\left(z_i+\frac{\delta z}{2},t\right).
   \label{coreeqn}
\end{eqnarray}
Here, the first term on the right-hand side represents the remaining electron density in the same state that survives any scattering event in the time interval past $[t, t+\delta t]$, and the second term denotes the effective source given in Eq.~\eqref{eq:seff}. The other four items correspond to the electron densities flowing into position $z_i$ from both sides through the interface $z_i\pm \delta z/2$. Note that the spin-dependent electron density is defined on the discrete grid $z_i$, while the interfaces are chosen in the middle of two neighboring grids, at which the interfacial reflectivity and the electron flux density are defined. \cite{battiato2014treating} For example, $\Phi_{\rightarrow}(z_i-\delta z/2,t)$ and $\Phi_{\leftarrow}(z_i+\delta z/2,t)$ indicate the electron flux densities flowing to the grid $z_i$ from its left and right sides, respectively. The electron flux density needs to be recursively solved,~\cite{battiato2014treating}
\begin{eqnarray}
\Phi_{\rightarrow}\left(z_i+\frac{\delta z}{2},t\right) &=& \sum_{t_0=0}^{t} S^{\rm eff}\left(z_i,t_0\right) \psi \left(\left. z_i+\frac{\delta z}{2},t\right\vert z_i,t_0\right)\nonumber\\
&&+\left[1-R_{\rightarrow}\left(z_i-\frac{\delta z}{2}\right)\right]\Phi_{\rightarrow}\left(z_i-\frac{\delta z}{2},t\right)+R_{\leftarrow}\left(z_i-\frac{\delta z}{2}\right)\Phi_{\leftarrow}\left(z_{i}-\frac{\delta z}{2},t\right),\nonumber\\
 \Phi_{\leftarrow}\left(z_i+\frac{\delta z}{2},t\right) &=& \sum_{t_0=0}^{t} S^{\rm eff}\left(z_{i+1},t_0\right) \psi \left(\left. z_i+\frac{\delta z}{2},t\right\vert z_{i+1},t_0\right)\nonumber\\
&&+\left[1-R_{\leftarrow}\left(z_{i+1}+\frac{\delta z}{2}\right)\right]\Phi_{\leftarrow}\left(z_{i+1}+\frac{\delta z}{2},t\right)+R_{\rightarrow}\left(z_{i+1}+\frac{\delta z}{2}\right)\Phi_{\rightarrow}\left(z_{i+1}+\frac{\delta z}{2},t\right).
 \label{eq:flux}
\end{eqnarray}
The last two terms on the right-hand sides of the above equations represent the contributions of the incoming electron flux passing through the interface and the reflected outgoing flux for the same interface. Therefore Eq.~\eqref{eq:flux} can recursively compute all the flux densities for the whole space in both directions. In Eq.~\eqref{eq:flux}, we introduce an auxiliary integrated flux kernel function $\psi$ [with $\phi$ defined in Eq.~\eqref{eq:kernel}],
\begin{equation}
\psi\left(\left. z_i\pm \frac{\delta z}{2}, t \right\vert z_{0}, t_{0}\right)=\int_{t}^{t+\delta t} d t'\ e^{\frac{t'-t-\delta t}{\tau(z)}}\phi\left(\left.z \pm \frac{\delta z}{2}, t' \right\vert z_{0}, t_{0}\right).
\end{equation}

It is worth emphasizing that we only explicitly include the interfacial reflectivity $R_{\rightleftarrows}$ at the FM$|$NM interface. However, the electron reflection due to scattering occurs everywhere. Inside the same material, the electron reflection is implicitly taken into account by the finite lifetime and the fact that scattering redistributes the energy and momentum of the hot electrons. Here the collision-induced spin flip, which can be a consequence of Elliott-Yafet phonon scattering or an inelastic magnon scattering, is neglected because of their relatively small contribution.~\cite{carva2011ab,essert2011electron,battiato2012theory} 

In the practical calculation, we need to iteratively solve Eq.~\eqref{eq:seff}, Eq.~\eqref{coreeqn} and Eq.~\eqref{eq:flux}. Having the electron density and flux density, we are able to obtain the time-dependent magnetization ($A$ is the cross-sectional area)
\begin{equation}
M(t)=A\int dz\,m(z,t)=2\mu_B A\int dz\left\{\int dE\,\left[n_{\uparrow}(E,z,t)-n_{\downarrow}(E,z,t)\right]+n_{\uparrow}^{\rm occ}(z,t)-n_{\downarrow}^{\rm occ}(z,t)\right\}.
\end{equation}
Here, $n_{\sigma}^{\rm occ}$ is the electron density below the Fermi level with spin $\sigma$, and it includes the equilibrium spin-dependent density $n^{\rm eq}_{\sigma}$ and the variation due to excitation and decay,
\begin{equation}
n^{\rm occ}_{\sigma}(z,t)=n^{\rm eq}_{\sigma}-\int_{-\infty}^t dt'\left\{\int_{E\ge E_F} dE\, S^{\rm ext}_{\sigma}(E,z,t')-\sum_{\sigma'}\int_{E<E_F} dE\int_{E'\ge E_F} dE'\,p_{\sigma,\sigma'}(E,E',z)\frac{n_{\sigma}\left(E',z,t'\right)}{\tau_{\sigma'}(E',z)}\right\}.
\end{equation}
The second term on the right-hand side represents the decrease in occupied electrons owing to the laser excitation, and the last term indicates the increased occupation below $E_F$ due to the hot-electron decay. Note that $M_s=2\mu_B AD(n_{\uparrow}^{\rm eq}-n_{\downarrow}^{\rm eq})$, where $D$ is the thickness of the FM metal.

The occupied electrons have relatively low mobility compared to the excited electrons above $E_F$, and therefore, their contribution to the spin current is neglected in the superdiffusive spin transport model. The spin current density is defined as the difference between spin-up and spin-down electron flux densities,
\begin{equation}
j_{s}(E,z,t) =\frac{\hbar}{2}\left[\Phi_{\rightarrow}^{\uparrow}\left(E,z,t\right)-\Phi_{\rightarrow}^{\downarrow}\left(E,z,t\right)-\Phi_{\leftarrow}^{\uparrow}\left(E,z,t\right)+\Phi_{\leftarrow}^{\downarrow}\left(E,z,t\right)\right].
\end{equation}

\end{widetext}

The excitation laser pulse is modeled using a Gaussian function with a wavelength of 780~nm (1.5~eV) and a full width at half maximum (FWHM) of 40~fs (60~fs) when calculating the ultrafast demagnetization of Fe (Ni). For Ni, the intensity of the pumped laser is taken as $8$~mJ$/{\rm cm}^2$, and 30\% of the light is assumed to be absorbed, \cite{stamm2007femtosecond} corresponding to 0.07 electrons excited per Ni atom. The ratio between the excited spin-up and spin-down electrons is 0.366:0.634. \cite{oppeneer2004ultrafast} The computational parameters for Fe are the same as in the previous calculation. \cite{battiato2012theory} The spin-dependent inelastic lifetimes and velocities are taken from first-principles many-body calculations~\cite{zhukov2005gw+, zhukov2006lifetimes}. The transition probability function $p_{\sigma,\sigma'}(E,E',z)$ is set to be the same as that in the literature \cite{battiato2012theory}. Numerically, we use a spatial grid of $\delta z=0.5$~nm and a time step of $\delta t=1$~fs. The energy range from $E_F$ to $E_F+1.5$~eV is considered in the calculation, which is discretized with an energy interval of $\delta E=0.125$~eV.

It is important to note that the spin-phonon relaxation time and electron-phonon relaxation time are both of the order of several picoseconds~\cite{koopmans2010explaining,bigot2000electron} corresponding to the increase of the lattice temperature. In this paper, we only calculate the dynamics within several hundreds of femtoseconds corresponding to the process of electron thermalization. Therefore, our treatment with neglecting phonons is valid unless the magnetization dynamics on a larger timescale needs to be studied.

\subsection{Ultrafast demagnetization in Fe}

\begin{figure}[t]
  \centering
  \includegraphics[width=\columnwidth]{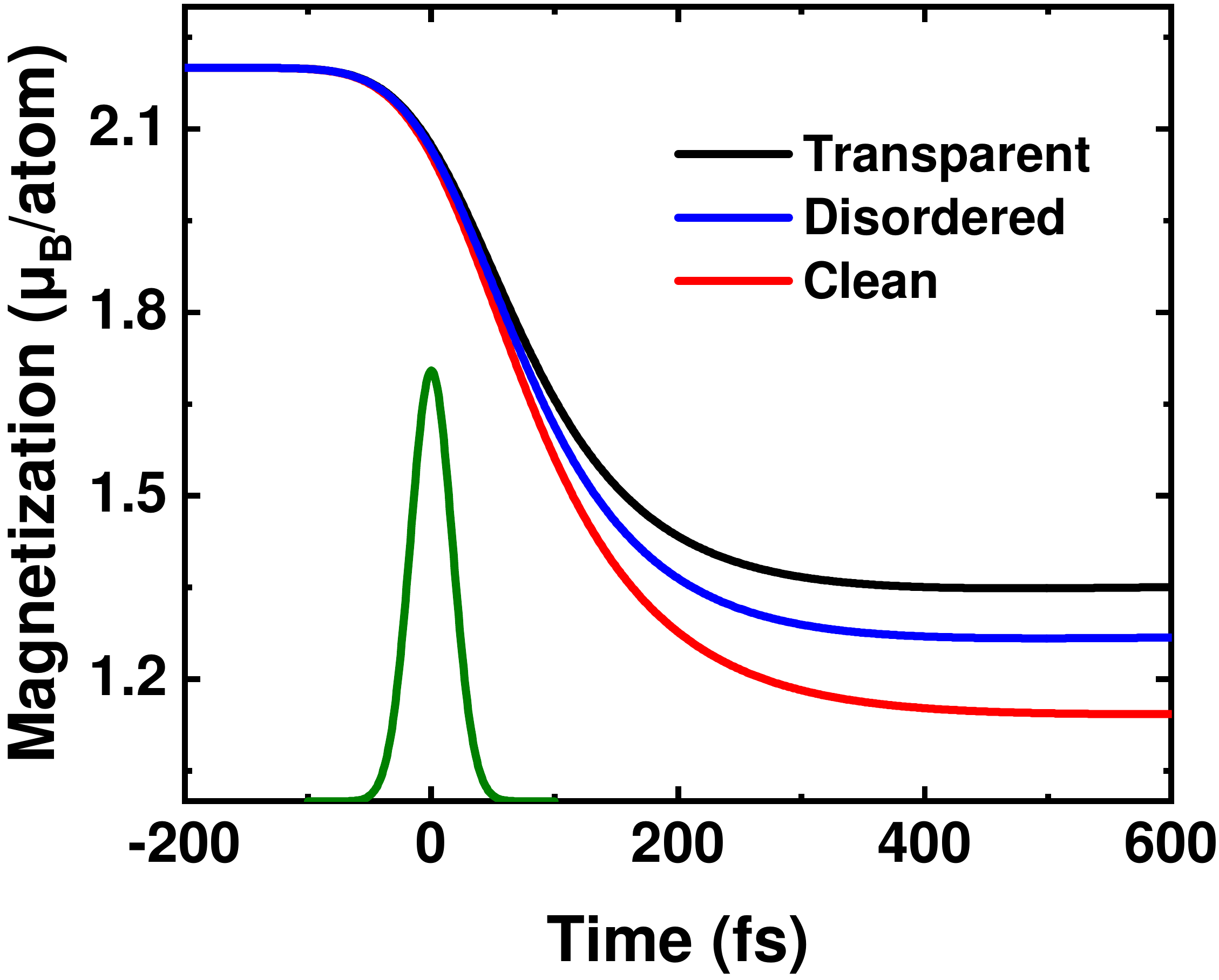}
  \caption{Calculated laser-induced demagnetization of a 10-nm-thick Fe film in contact with a semi-infinite Au layer. The red and blue lines correspond to the calculations using the reflectivity at clean and disordered Fe$|$Au interfaces, as shown in Fig.~\ref{fig:R-FeNM}. The green line illustrates the laser pulse. The demagnetization curve for a transparent Fe$|$Au interface is also plotted using a black line for comparison, where all the excited carriers can freely pass through the interface from both sides.}
  \label{fig:FeAu}
\end{figure}
Using the superdiffusive spin transport model, we calculate ultrafast demagnetization in an Fe(10 nm)$|$Au bilayer and plot the magnetization as a function of time in Fig.~\ref{fig:FeAu}. To model the infinitely thick Au layer, we consider a 40-nm-thick Au film in the practical calculation and set the reflection at its right boundary to be zero. A Gaussian laser pulse with FWHM$=$40~fs is applied as the excitation, as plotted using a green line in Fig.~\ref{fig:FeAu}. Following the excitation, the magnetization of Fe decreases very rapidly to approximately 50\% of its equilibrium $M_s$.

Compared with the transparent interface used in the literature, the reflectivity of the real interface included in the calculation results in a larger decrease in the ultrafast demagnetization. In the clean interface case, the decrease in magnetization saturates at approximately 1.1~$\mu_B$ per Fe atom, which is approximately 15\% smaller than the value obtained using the transparent interface. The stronger demagnetization occurs because of the spin-dependent interface resistance shown in Fig.~\ref{fig:FeAu-conductance}: the lower reflectivity of spin-up electrons leads to a higher efficiency in transferring the majority-spin angular momentum into the Au layer and therefore enhances the laser-induced demagnetization. The disordered Fe$|$Au interface has a weaker spin filter effect than the clean one such that the demagnetization strength is also slightly weaker.

\begin{figure}[b]
  \centering
  \includegraphics[width=\columnwidth]{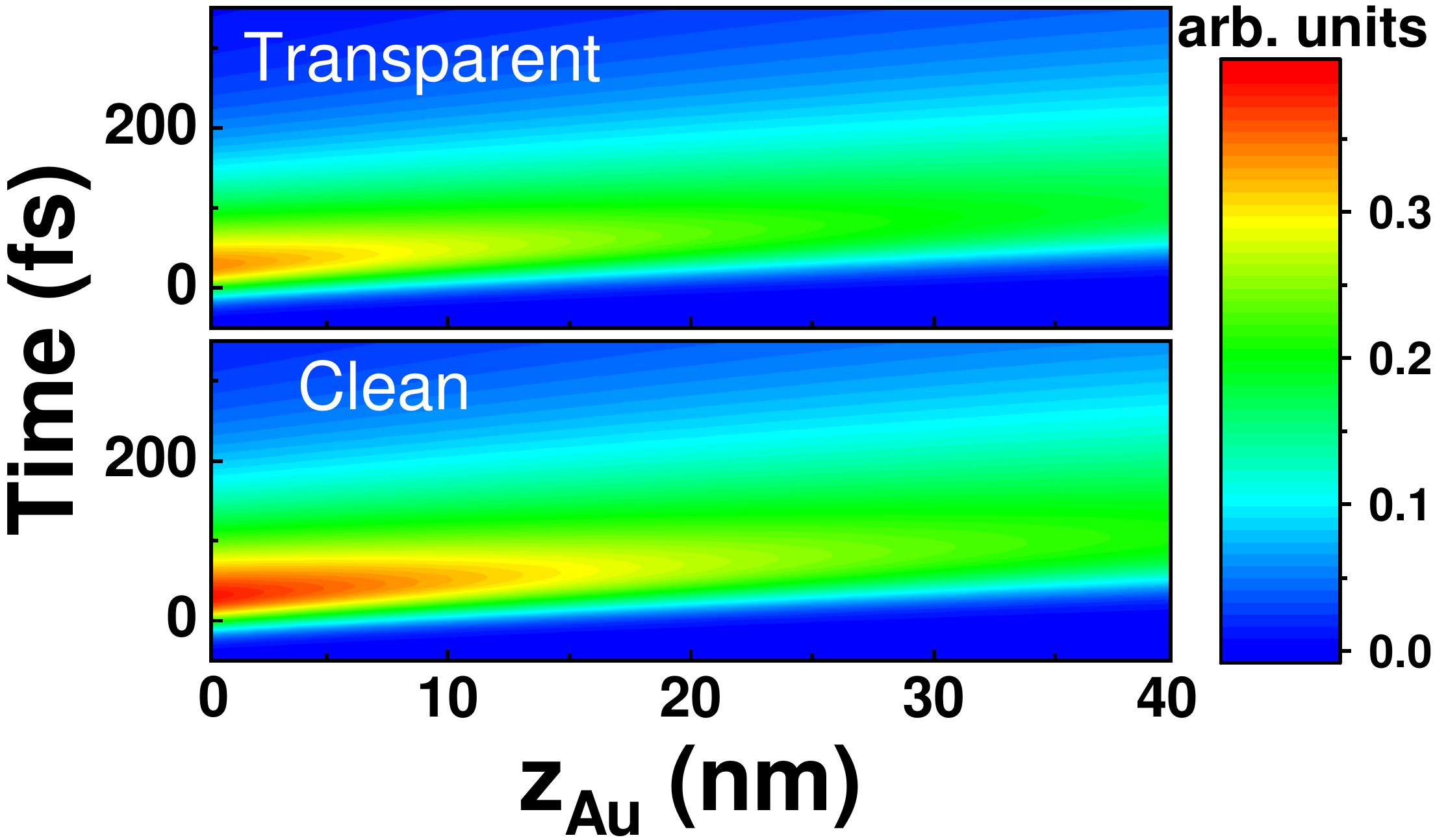}
  \caption{Calculated spin current in Au as a function of position and time. The upper and lower panels correspond to the transparent and real (clean) Fe$|$Au interfaces used in the superdiffusive spin transport model, respectively.}
  \label{fig:spincAu}
\end{figure}
\begin{figure}[t]
  \centering
  \includegraphics[width=\columnwidth]{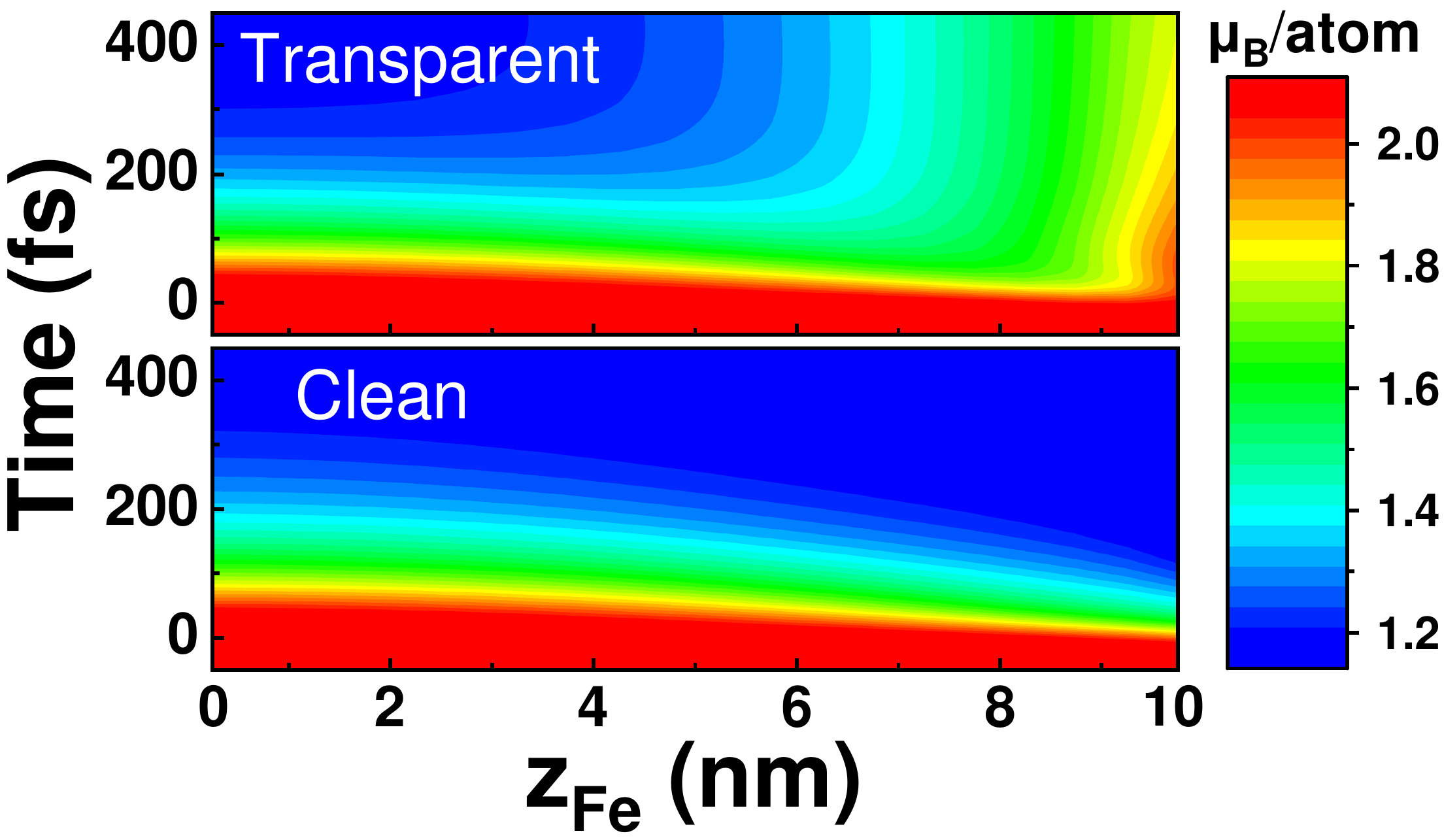}
  \caption{Calculated distribution of magnetization in the Fe layer. The upper and lower panels correspond to the transparent and real (clean) Fe$|$Au interfaces used in the superdiffusive spin transport model, respectively. }
  \label{fig:mFe}
\end{figure}
The spin current injected into Au is shown in Fig.~\ref{fig:spincAu} for both transparent and clean Fe$|$Au interfaces. Owing to the spin filter effect, the spin-up electrons move more easily across the interface, and the resulting spin current, which is injected into Au, is larger than that for the transparent interface. This is in agreement with the stronger demagnetization in Fig.~\ref{fig:FeAu}.

Figure~\ref{fig:mFe} shows the local magnetization variation in the Fe layer after the laser excitation for the transparent and clean Fe$|$Au interfaces. Due to the application of the laser pulse, the whole Fe film exhibits significant demagnetization around $t=100$~fs for both the transparent and clean interfaces. The magnetization decrease is stronger on the right side close to the Au layer because, at the Fe$|$Au interface, the spin can be transferred into Au, while the left surface of Fe reflects the spins carried by hot electrons. It is interesting to note that spin accumulates at the transparent interface in the upper panel of Fig.~\ref{fig:mFe}. Nevertheless, this spin accumulation disappears in the calculation with the real Fe$|$Au interface; see the lower panel. Such a difference is responsible for the difference in the ultrafast demagnetization shown in Fig.~\ref{fig:FeAu}.

\begin{figure}[t]
  \centering
  \includegraphics[width=\columnwidth]{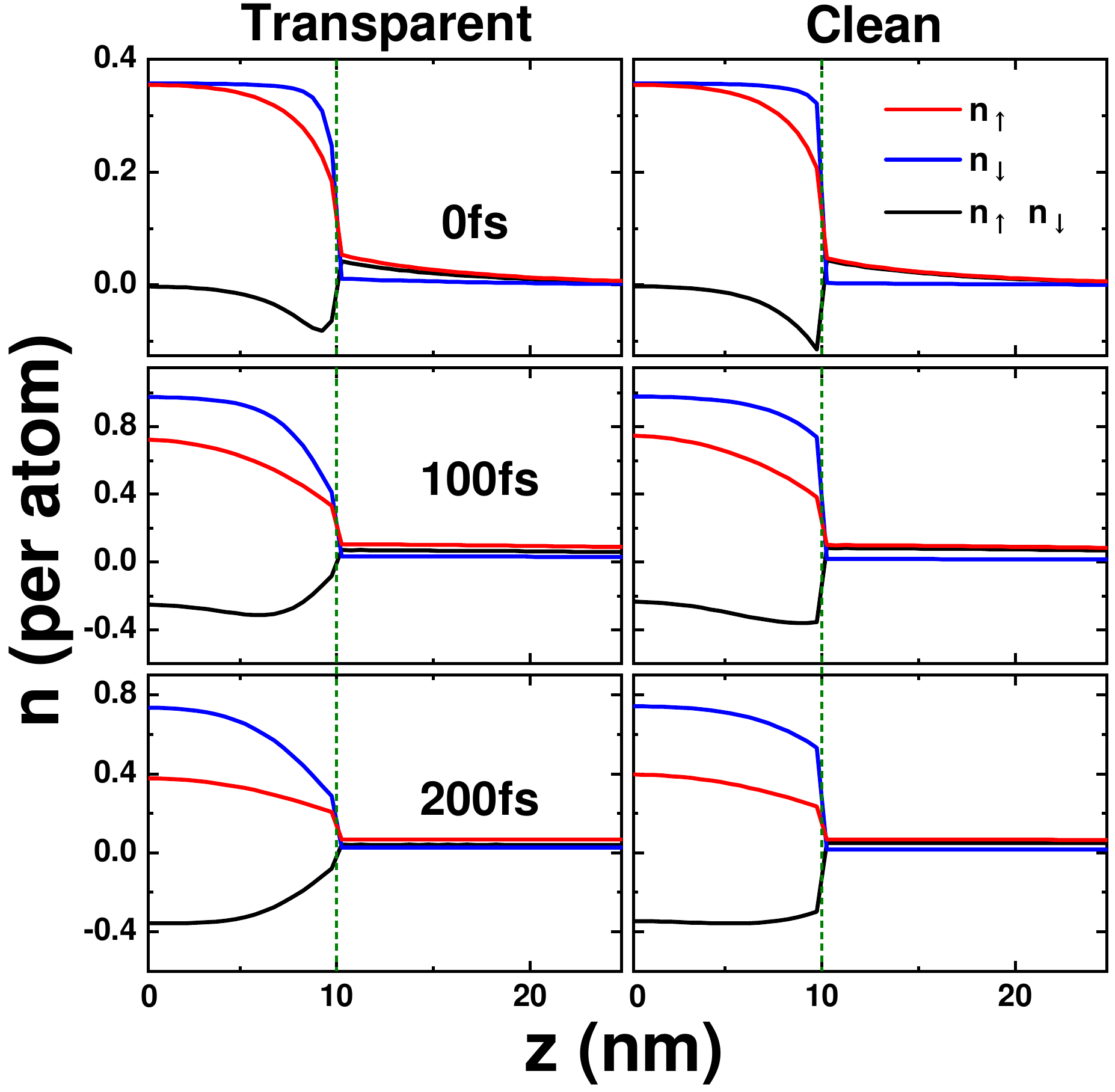}
  \caption{Calculated spin-dependent electron density above the Fermi level for transparent (left panels) and real clean (right panels) Fe$|$Au interfaces. The black lines indicate the difference between spin-up and spin-down electrons. The vertical dashed lines illustrate the interface position.}
  \label{fig:spindensityFeAu}
\end{figure}
To understand the differences in the interfacial spin accumulation between the transparent and real interfacial reflections, we plot in Fig.~\ref{fig:spindensityFeAu} the excited spin-dependent electron densities in the Fe$|$Au bilayer. The laser pulse excites equal numbers of spin-up and spin-down electrons to energies above the Fermi level in Fe. Because the velocity of spin-up electrons is larger, they exhibit more efficient transport into the Au layer. Thus, the spin-up electron density of Fe (the red lines in Fig.~\ref{fig:spindensityFeAu}) becomes lower than the spin-down electron density (the blue lines). Then, the loss of the majority-spin (spin-up) electrons from the Fe layer results in the demagnetization seen in Fig.~\ref{fig:FeAu}. With a transparent interface, the high density of spin-down electrons in Fe drives a strong diffusion across the interface into Au, which is more significant than the diffusion of the spin-up electrons. Such a leakage of spin-down electrons leads to the accumulation of spin-up electrons shown in the upper panel of Fig.~\ref{fig:mFe}. However, a real Fe$|$Au interface has a very high reflectivity for the spin-down electrons. Although the spin-down electron density in Fe near the interface is high, diffusion of these electrons into Au is difficult. On the other hand, the low reflectivity of the interface helps spin-up electrons leave the Fe, removing the spurious spin accumulation due to the transparent interface, as shown in the lower panel of Fig.~\ref{fig:mFe}.

\begin{figure}[t]
  \centering
  \includegraphics[width=\columnwidth]{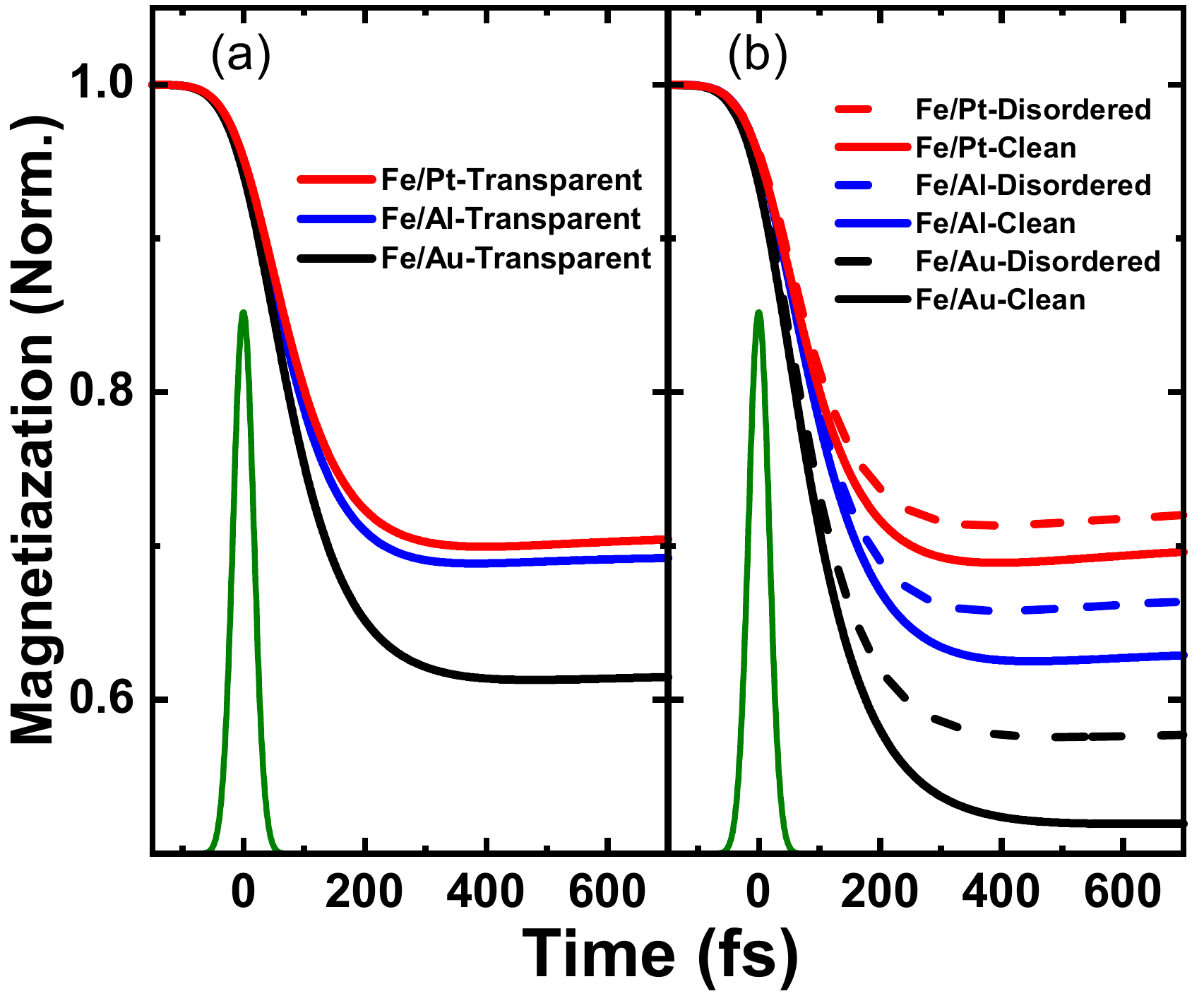}
  \caption{(a) Calculated demagnetization of Fe in the Fe$|$NM (NM=Au, Al, and Pt) bilayers with a transparent interface. (b) Calculated demagnetization with the real reflectivities of the clean and disordered interfaces. The solid and dashed lines are obtained using the clean and disordered interfacial reflectivities, respectively. The green lines illustrate the profile of the laser pulse.}
  \label{fig:FeNMdem}
\end{figure}
Figure~\ref{fig:FeNMdem}(a) shows the calculated demagnetization of a 10-nm-thick Fe film in contact with infinitely thick Au, Al, and Pt layers, where the interface is assumed to be transparent. The Fe$|$Au bilayer exhibits the strongest demagnetization, while the demagnetization is comparable in the Fe$|$Al and Fe$|$Pt systems. The main factors that determine such differences are the mobility and lifetime of the hot electrons in the NM metal. The hot electrons in Au have the largest group velocity and thus efficiently transfer the spin angular momentum away from the interface, resulting in the ultrafast demagnetization of Fe.\cite{} The longer lifetime of the hot electrons in Au reduces the electron scattering rate and prevents the injection of spin current back into Fe.

We plot the demagnetization of the Fe film using the real interfacial reflectivity in Fig.~\ref{fig:FeNMdem}(b). Unlike the transparent interface, there is a noticeable difference between the Fe$|$Al and Fe$|$Pt bilayers. This difference arises because the high reflectivity from Al to Fe [see Fig.~\ref{fig:R-FeNM}(d)] strongly suppresses the spin backflow and hence enhances the demagnetization in Fe. For all three NM metals, the spin filter effect is more significant at the clean interface than at the disordered one, and the decrease in magnetization is therefore stronger with the clean Fe$|$NM interface. The above numerical results suggest that the interfacial reflectivity plays an important role in the theoretical description of laser-induced ultrafast demagnetization and should not be neglected. The influence of the thickness of the disordered layers is examined using the Fe$|$Au bilayer as an example, where the difference in the calculated results is less than 2\% for two, four and six atomic layers of the disordered alloy interface.


\subsection{Ultrafast demagnetization in Ni}

\begin{figure}[t]
  \centering
  \includegraphics[width=\columnwidth]{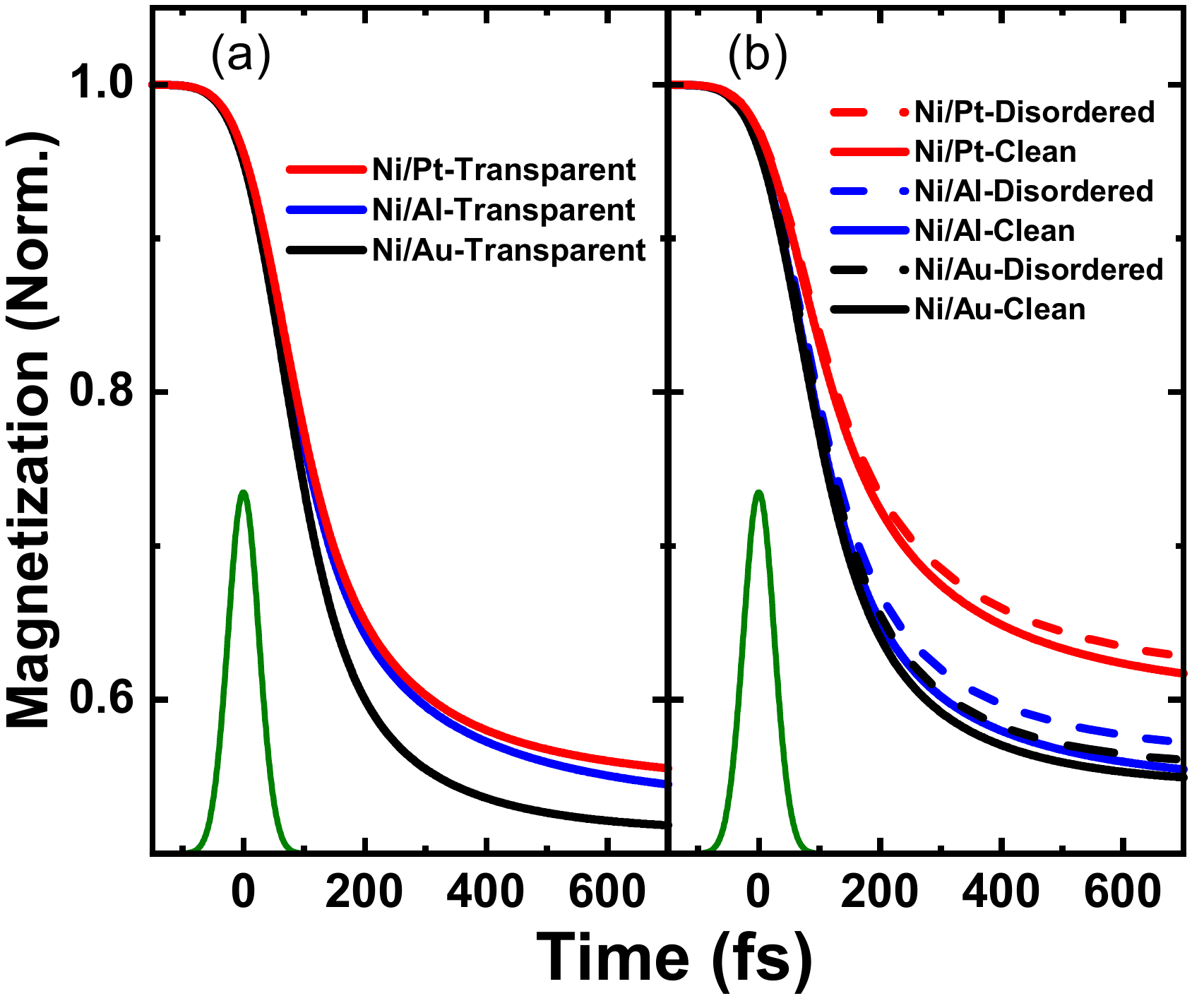}
  \caption{(a) Calculated demagnetization of Ni in the Ni$|$NM (NM=Au, Al, and Pt) bilayers with a transparent interface. (b) Calculated demagnetization with the real reflectivities of clean and disordered interfaces. The solid and dashed lines are obtained using the clean and disordered interfacial reflectivities, respectively. The green lines illustrate the profile of the laser pulse. The thickness of the Ni film is 15~nm, and the FWHM of the laser pulse is 60~fs.}
  \label{fig:NiNMdem}
\end{figure}

Then, we consider the ultrafast demagnetization of a nickel film of thickness 15~nm that is on top of semi-infinite Au, Al, and Pt layers. Using a transparent interface, the calculated demagnetization is shown in Fig.~\ref{fig:NiNMdem}(a). Since the FWHM of the laser is 60~fs, the decrease in magnetization is slower than that in Fig.~\ref{fig:FeNMdem}(a). The normalized demagnetization strength is larger for Ni than Fe because the excited hot electrons in Ni are mostly $s$ electrons, except for the spin-down channel near the Fermi level. Then, the hot electrons move faster in Ni than those in Fe and can more efficiently transfer the spin angular momentum into the adjacent NM layer. Consistent with the analysis for the Fe$|$NM cases, we find that the Ni$|$Al and Ni$|$Pt bilayers are comparable, while the Ni$|$Au bilayer has the strongest demagnetization owing to the large mobility and long lifetime of hot electrons in Au.

When taking the real interface reflectivity into account, we see that the demagnetization of Ni$|$NM becomes weaker for all three NM metals, in sharp contrast to the cases of Fe$|$NM. The interfacial reflectivities shown in Fig.~\ref{fig:R-NiNM} are nearly spin independent for the Ni$|$NM systems, unlike the spin filtering at the Fe$|$NM interfaces. Compared with the transparent interface, the real interface simultaneously decreases the transmission of spin-up and spin-down electrons. Thus, the efficiency of transferring spin angular momentum from Ni to the NM metals is weaker with the real interfacial reflectivity. As seen in Fig.~\ref{fig:R-NiNM}, the alloy disorder at the Ni$|$NM interface only plays a minor role in its reflectivity. Thus, the calculated demagnetization of the Ni$|$NM bilayer is comparable between the clean and disordered interfaces. In addition, the real Ni$|$Al interface exhibits a substantial reflectivity from Al to Ni, which remarkably suppresses flow of the transferred spin current in Al back into Ni. Therefore, the demagnetization of the Ni$|$Al bilayer is substantially stronger than that in the Ni$|$Pt system, although they are comparable in the presence of a transparent interface.

\section{\label{sec:level4}THz emission}\label{chap4}

\begin{figure}[t]
  \centering
  \includegraphics[width=\columnwidth]{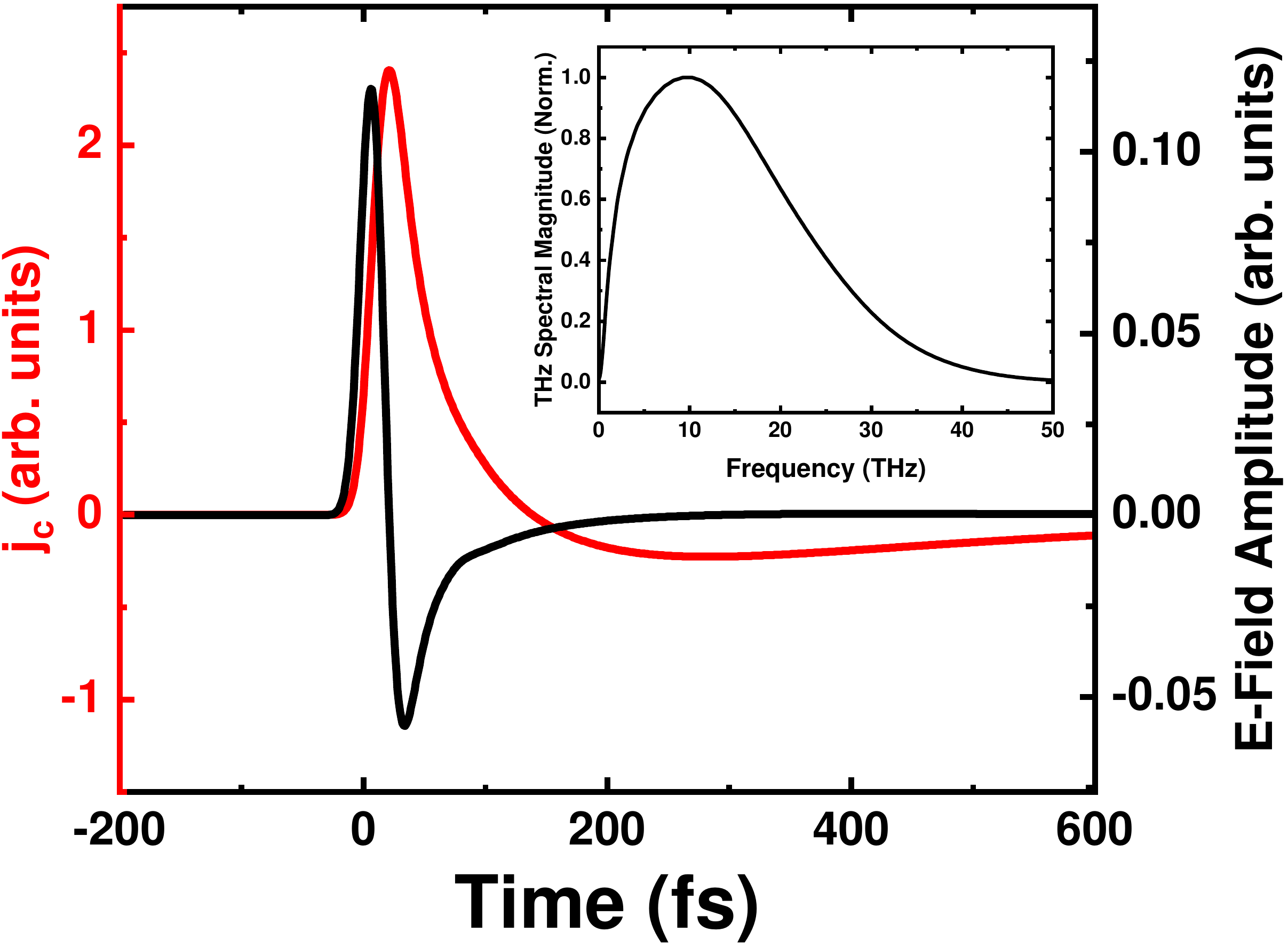}
  \caption{Calculated electric current density in Au as a function of time (red line). The Fe(10~nm)$|$Au(15~nm) bilayer is excited by a 40-fs-wide laser pulse. The black line shows the emitted electric field amplitude in the far-field approximation. Inset: Fourier transform amplitude of the emitted electric field as a function of frequency.}
  \label{fig:THz1}
\end{figure}

The spin current carried by the laser-driven hot electrons flowing into the NM metal can generate a transverse charge current via the ISHE~\cite{Hoffmann:2013,Niimi:2015,Sinova:2015}
\be
\mathbf j _c = \Theta_{\rm sH} \left(-\frac{2e}{\hbar}\right)\bm{\sigma}\times\mathbf j_s,
\ee
where $\mathbf j_c$ and $\mathbf j_s$ are the charge and spin current densities, respectively, and $\bm{\sigma}$ represents the polarization of $\mathbf j_s$. The charge-spin conversion efficiency is given by the dimensionless parameter $\Theta_{\rm sH}$, which is usually called the spin Hall angle. Since $\mathbf j_c$ varies in time, the time-dependent charge current generates electromagnetic waves. In the far-field approximation, the amplitude of the electric field is determined by the time derivative of $\mathbf j_c$ \cite{kuvzel1999spatiotemporal},
\be
\mathbf E=\frac{w_0^2}{c L} Z \int dz\,\frac{\partial\mathbf j_c(z)}{\partial t},
\ee
where $Z$ is the impedance of the bilayer including the contributions from the half space of air and the other half space of the substrate\cite{seifert2016efficient}, $w_0$ is the terahertz beam waist and here refers to the radius of the laser pulse spot, $L$ denotes the distance between the terahertz emitter and the detector, and $c$ is the speed of light.

The above computational scheme is illustrated using a Fe(10~nm)$|$Au(15~nm) bilayer. The red line in Fig.~\ref{fig:THz1} shows the average electric current density in Au as a function of time. Here, we only consider the relative strength such that the particular value of the spin Hall angle does not play a role. The electric current pulse exhibits an asymmetric shape: its quick increase at $t<$~21 fs results in a large positive peak in the emitted electric field (the black line), while the slow decay at $t>$~21 fs gives rise to the smaller negative peak. The width of the electric current pulse is approximately 100~fs, corresponding to the terahertz frequency regime. The spectrum in the frequency domain is obtained by Fourier transform of the emitted electric field, which is plotted in the inset of Fig.~\ref{fig:THz1}.

\begin{figure}[t]
  \centering
  \includegraphics[width=\columnwidth]{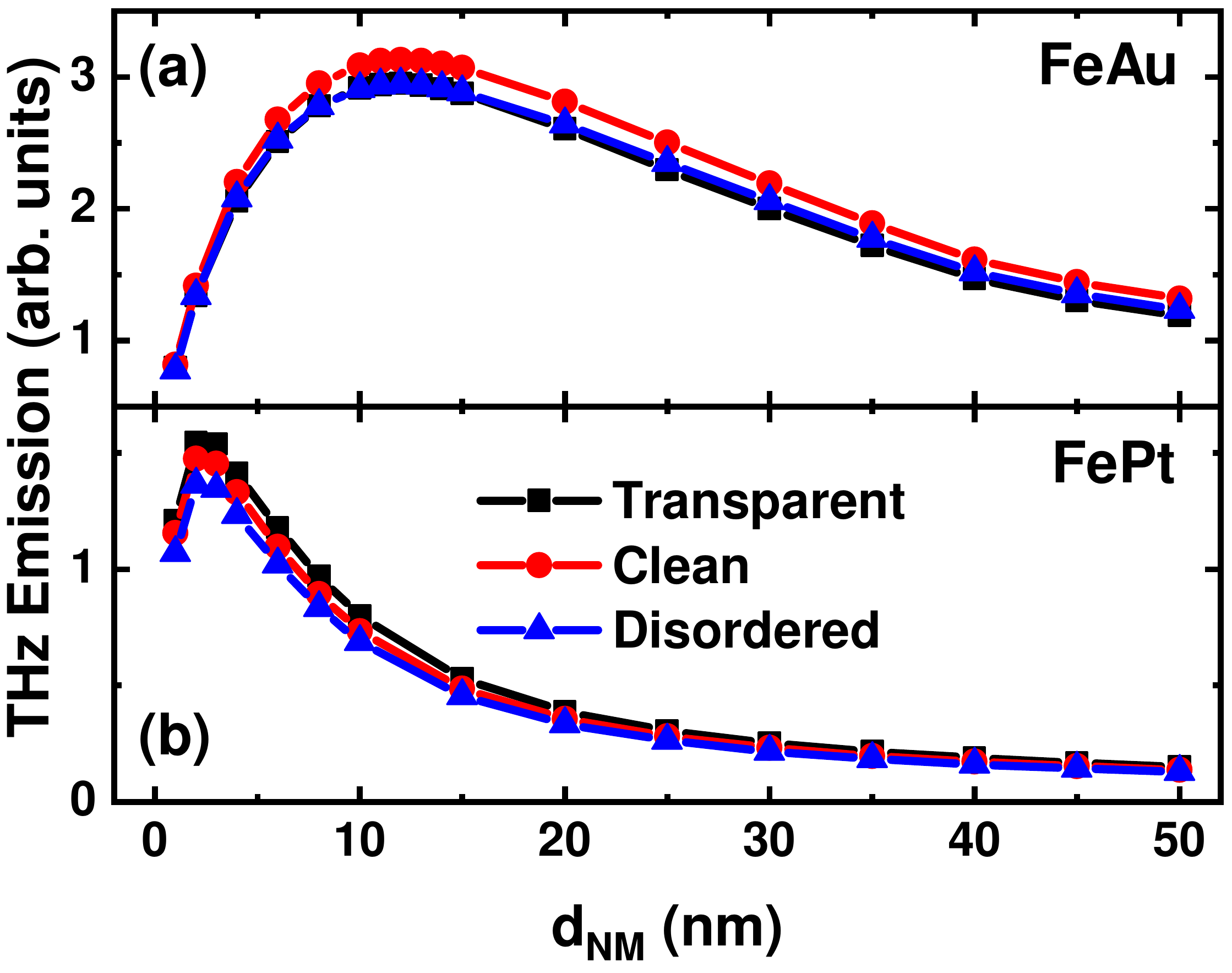}
  \caption{Calculated THz emission intensity, which is measured by the difference in the maximum and minimum $E$-field amplitudes, as a function of the Au thickness. The Fe film is 10 nm thick. The transparent and real interfaces, including the clean and disordered ones, are considered in the calculations.}
  \label{fig:THz-Fe}
\end{figure}
In the following, we use the difference between the maximum and minimum $E$-field amplitudes as a measure of the THz emission strength, which is plotted in Fig.~\ref{fig:THz-Fe}(a) as a function of the Au thickness. In general, the THz emission shows a non-monotonic dependence on the Au thickness $d_{\rm Au}$. At small $d_{\rm Au}$, the injected spin current is reflected back to Fe before it is fully exploited to generate a transverse charge current such that the THz emission increases with increasing  $d_{\rm Au}$. When the Au layer is thick enough, the total transverse charge current is expected to saturate, while the decrease in the impedance $Z$ leads to a decrease in the emitted THz amplitude.~\cite{seifert2016efficient} Because the real Fe$|$Au interface has a spin filtering effect, including the reflectivity of the real interface increases the spin current injected into Au. Therefore, the calculated THz emission is larger for the clean interface than for the transparent one, as shown in Fig.~\ref{fig:THz-Fe}(a). This indicates that the THz emission can be optimized by an effective spin filter at the interface.

Figure~\ref{fig:THz-Fe}(b) shows the calculated THz emission of the Fe$|$Pt bilayer as a function of the Pt thickness, which also exhibits a non-monotonic behavior. The maximum value is located at $d_{\rm Pt}\approx 3$~nm and is much smaller than that in the case of Fe$|$Au. This occurs because the spin diffusion length of Pt is much shorter than that of Au such that the spin current in Pt is fully converted to the transverse charge current within a very short length scale. It is interesting to note that the emitted electromagnetic field amplitude is smaller for the real interface reflectivity, in contrast to Fe$|$Au. Although the real Fe$|$Pt interface reflectivity increases the spin polarization of the current injected into Pt compared to the transparent interface, it decreases the absolute amplitude of the current. Therefore, to increase the THz emission, one should maximize both the amplitude and spin polarization of the current injected into the NM metal.

\begin{figure}[t]
  \centering
  \includegraphics[width=\columnwidth]{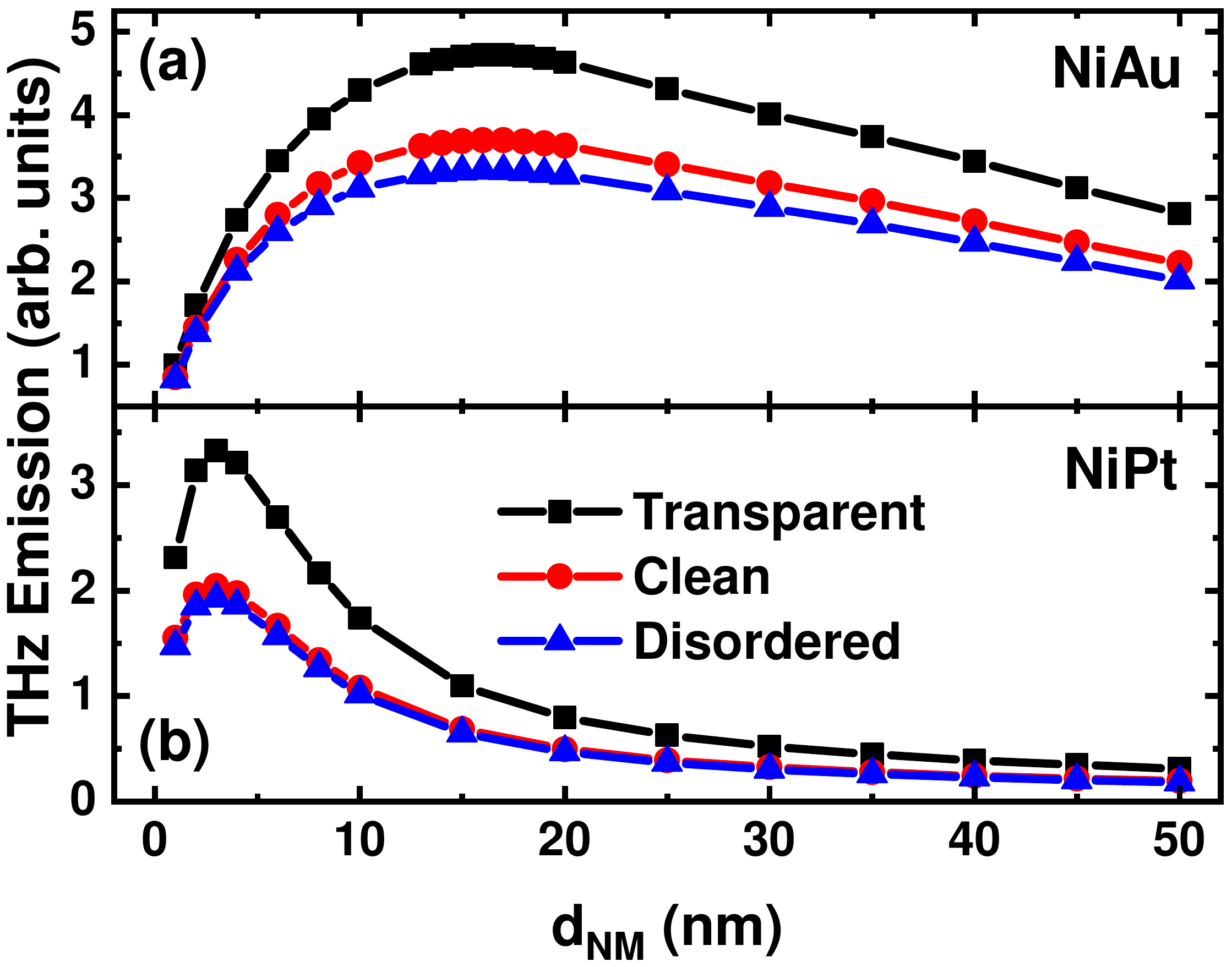}
  \caption{Calculated THz emission intensity, which is measured by the difference in the maximum and minimum $E$-field amplitudes, as a function of the Ni thickness. The Ni film is 10 nm thick. The transparent interface and the real interfaces, including the clean and disordered ones, are considered in the calculations.}
  \label{fig:THz-Ni}
\end{figure}

The calculated THz emission amplitudes of the Ni$|$Au and Ni$|$Pt bilayers are plotted in Fig.~\ref{fig:THz-Ni}. The thickness of the Ni film is 10 nm. Since the Ni$|$Au and Ni$|$Pt interfaces do not have a spin filter effect, the real interface reflectivity only decreases the total amount of spin current injected into the NM metal, resulting in a weaker emitted electromagnetic field compared with the corresponding transparent interface. In analogy with the Fe$|$NM bilayers, the THz emission by the Ni$|$NM systems also shows a non-monotonic dependence on the NM thickness. Moreover, the maximum value occurs at a shorter thickness for the Ni$|$Pt system than for the Ni$|$Au system owing to the shorter spin diffusion length of Pt.

In addition to the transverse charge current generated via the ISHE, a tilting magnetization can also generate a charge current at an interface as the inverse effect of spin-orbit torque. In a recent experiment, helicity-dependent in-plane photocurrents are observed at FM$|$NM interfaces due to the Rashba spin-orbit interaction~\cite{huisman2016femtosecond}. This effect results in the THz field parallel to the magnetization of the FM metal while the ISHE leads to the field perpendicular to the magnetization. Moreover, the THz field strength arising from interface Rashba effect is much smaller than the contribution from ISHE and therefore is neglected in this paper.

\section{\label{sec:level5}Conclusions and discussions}\label{chap5}

We have calculated the spin- and energy-dependent electron reflectivity at the FM$|$NM interfaces with FM=Fe and Ni and NM=Au, Al, and Pt using first-principles scattering theory. The Fe$|$NM interfaces show a spin filtering effect owing to wave function mismatch between spin-dependent electronic states above the Fermi energy. The calculated interface reflectivity is incorporated into the superdiffusive spin transport theory, which is used to study the laser-induced demagnetization and THz emission in Fe$|$NM and Ni$|$NM heterostructures. The spin filtering effects at the Fe$|$NM interfaces enhance the demagnetization and electromagnetic radiation amplitude. Using Fe$|$Au as an example, we demonstrate that the spurious spin accumulation in Fe close to the Fe$|$Au interface is eliminated by employing the calculated reflectivity of the real interface. For the Ni$|$NM bilayers, the calculated demagnetization and THz emission are weaker for the real interface reflectivity because the amplitude of the spin current injected into the NM metal is smaller. These results suggest that the reflectivity of real interfaces has to be taken into account in a quantitative description of laser-induced demagnetization and THz emission.

\begin{figure}[t]
  \centering
  \includegraphics[width=0.9\columnwidth]{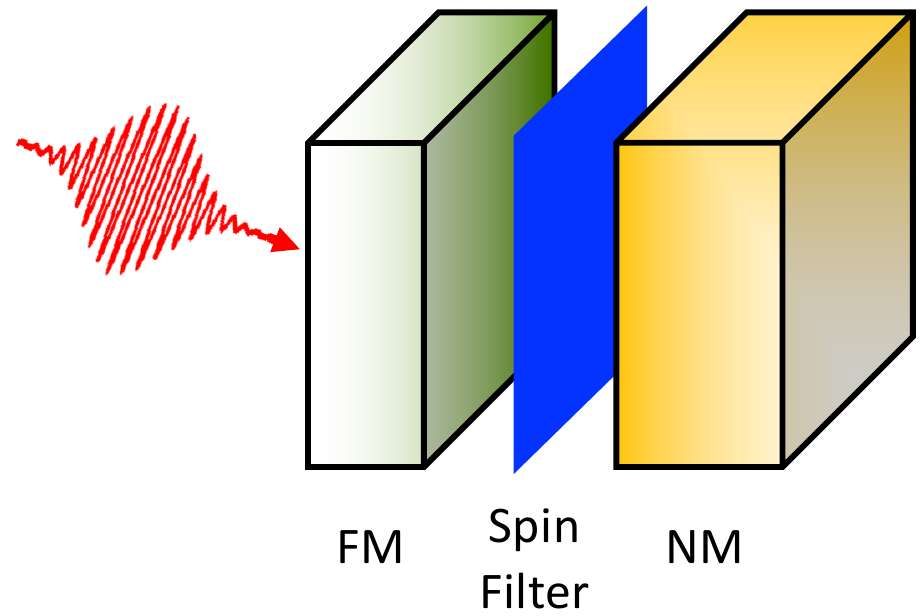}
  \caption{Proposed THz emitter consisting of multilayers. The left FM metal can be chosen to have a high efficiency for femtosecond laser absorption so as to maximize the excited hot electron density. The middle layer is an effective spin filter, which has a high reflectivity for one spin and a low reflectivity for the opposite spin. The right NM layer is a heavy metal with a large efficiency in converting the injected spin current into a transverse charge current.}
  \label{fig:proposal}
\end{figure}

Noting the important role of the interface reflection and transmission allows us to have more flexibility in designing and optimizing spintronic THz emitters. As schematically illustrated in Fig.~\ref{fig:proposal}, we propose an efficient spintronic emitter with a coated interface. The main functionality of the FM layer on the left side is to absorb the laser pulse and create as many hot electrons as possible. Therefore, a FM metal or an alloy with a high density of states above the Fermi energy is preferred. The spin filter between the left-hand-side metal and the right-hand-side NM metal can be an FM metal or an alloy, as well as some coating at the interface. This results in large transmission of one spin and small transmission of the opposite spin such that the hot electrons are highly spin-polarized after being injected into the NM metal, which should have a large spin-to-charge conversion efficiency, e.g., a large spin Hall angle, and the radiated THz electromagnetic wave can be maximized. For example, we expect that inserting a very thin Fe layer between Ni and the NM metal could significantly enhance the spin polarization of the hot electrons entering the NM metal. Therefore, the THz emission of the Ni$|$Fe$|$NM multilayer is expected to be stronger than that of the Ni$|$NM bilayer. 

\acknowledgements
We thank T. Kampfrath, C. Stamm, Ka Shen and Luo Zhang for helpful discussions and communications. This work was partly supported by the National Natural Science Foundation of China (Grants No. 61774018 and No. 11734004), the Recruitment Program of Global Youth Experts, and the Fundamental Research Funds for the Central Universities (Grants No. 2018EYT03 and No. 2018STUD03).

\end{document}